# Polymorphic phase transitions in bulk triglyceride mixtures

**Diana Cholakova, Slavka Tcholakova, Nikolai Denkov***

*Department of Chemical and Pharmaceutical Engineering*
*Faculty of Chemistry and Pharmacy, Sofia University,*
*1 James Bourchier Avenue, 1164 Sofia, Bulgaria*

*Corresponding authors:
Prof. Nikolai Denkov
Department of Chemical and Pharmaceutical Engineering
Sofia University
1 James Bourchier Ave.,
Sofia 1164
Bulgaria
E-mail: nd@lcpe.uni-sofia.bg
Tel: +359 2 8161639
Fax: +359 2 9625643

# ABSTRACT

Triacylglycerols (TAGs) are among the most important ingredients in food, cosmetic and pharmaceutical products. Many physical properties of such products, incl. morphology, texture and rheology, are determined by the phase behaviour of the included TAGs. Triglycerides are also of special interest for the production of solid lipid nanoparticles, applied for controlled drug delivery and for encapsulation of bioactive ingredients. In this paper, we study the polymorphic behaviour of complex TAG mixtures, composed of 2 to 6 mixed TAGs, by differential scanning calorimetry and X-ray scattering techniques, aiming to reveal the general rules for their phase behaviour upon cooling and heating. The results show that two or more coexisting phases form upon solidification (α, β' and/or β), the number of which depends strongly on the cooling rate and on the number of components in the mixture. No completely miscible α- or β'-phases were observed. The structure of the most stable β polymorphs, formed upon subsequent heating of the solidified samples, does not depend on the thermal history of the samples. For all mixtures studied, we observed one-component β domains, coexisting with binary mixed β domains with composition and structure which do not depend on the specific TAG ratio in the mixture. In other words, for a mixture with $k$ saturated TAGs we observed $(2k-1)$ different β phases. These conclusions provide some predictive power when analysing the phase transition properties of TAG mixtures.

## 1. Introduction

Triacylglycerols (TAG), also called triglycerides for simplicity, are esters of glycerol (propane-1,2,3-triol) and three fatty acids. They are widely used in pharmaceutical, food and cosmetic products as lipophilic matrices for dispersal or dissolution of essential ingredients, or as structuring components of the final products [1-9]. TAGs also play an indispensable role in the human diet, providing essential fatty acids and many vital lipophilic actives, and serving as an important energy source [5-8]. Depending on the type and chain length of the fatty acid residues in the TAG molecules, they are divided into two major groups – monoacid triglycerides which have identical fatty acid residues and mixed-acid triglycerides with different chain lengths [10]. The natural oils and fats, *e.g.* coconut oil, palm kernel oil, sunflower oil, peanut butter, cocoa butter, lard, *etc.*, usually consist of a multicomponent mixture of mixed-acid triglycerides [8,10,11].

The phase behavior of natural oils and fats plays a crucial role for the texture, solubility, plasticity, glossiness, firmness and many other properties of various consumer products, incl. ice-cream, chocolate, margarine, butter, whipped cream and others in the food industry, and lip balms, creams and lotions in cosmetics [10,12,13]. Therefore, a deep understanding of the phase behavior of these systems is of fundamental importance for controlling and predicting these properties, and for optimization of the fat-based formulations when using healthier and/or more sustainable raw materials.

Triglycerides exhibit monotropic polymorphism [10,14,15], *i.e.* there is a single polymorphic phase (denoted as β), whose energy is the lowest in the whole temperature range in which it may exist, see Figure 1a,b. However, two other, less stable polymorphs, denoted as α and β', are known to form in simple monoacid triglycerides [10,14-16]. The stability and melting temperature of the various polymorphs increase in the order α→β'→β. Once any of the more stable phases is formed, the phase transition is irreversible, *i.e.* if a sample is arranged in β' phase and we want to obtain α phase, we need to melt the sample into the isotropic liquid phase and then cool it again. The polymorphic phase transitions may occur in two different ways: (1) upon heating *via* the so-called "melt-mediated" route, by melting of the less stable polymorph and subsequent recrystallization into the more stable one, or (2) upon prolonged storage at a temperature which is lower than the melting temperature of the respective unstable polymorph *via* the so-called "solid-state phase transition" [10,14-18].

The least stable α polymorph has hexagonally packed vertical chains which have almost statistical orientation [10,14], thus resembling in some aspects the rotator phases of alkanes

[19,20], although the chain mobility near the glycerol group in triglycerides is more restricted by steric constraints [21]. The intermediate in stability β' polymorph has an orthorhombic sub-cell with tilted chains, and the most stable β phase has triclinic parallel sub-cell with tilted chains [10,14], see Figure 1c. The melting temperatures of pure triglycerides in different polymorphic phases and their characteristic interplanar spacings have been extensively studied and are summarized in Figure 1d [10,14,15].

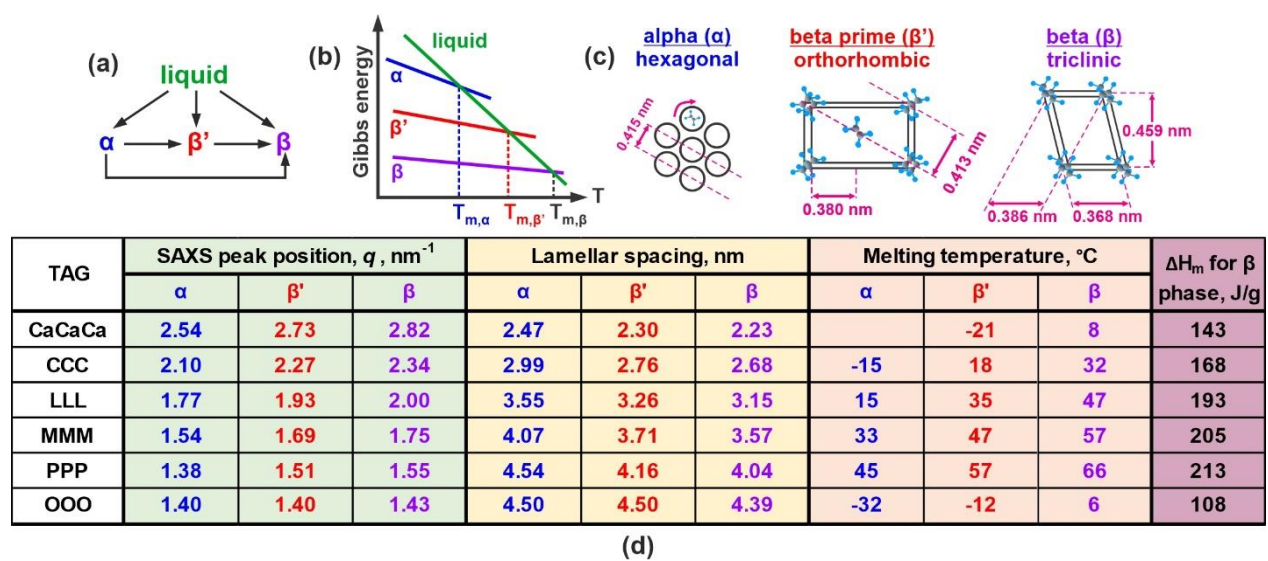

**Figure 1.** **Schematic presentation of the main TAG polymorphs and literature data about their properties.** (a) Possible phase transitions between liquid, α, β' and β phases in TAGs. (b) Schematic presentation of the free energy of the phases as a function of temperature. (c) Sub-cell structures of α, β' and β polymorphs of TAG crystals. Schematics in (a-c) are adapted from Refs. [10,15]. (d) Characteristic lamellar thicknesses, melting temperatures and enthalpies for pure triglycerides: tricaprylin (CaCaCa), tricaprin (CCC), trilaurin (LLL), trimyristin (MMM), tripalmitin (PPP) and triolein (OOO). SAXS positions for LLL and MMM are determined in the current study, whereas the other data are taken from Refs. [14,22].

The crystallization of TAG mixtures has been studied mainly in the context of natural oils and fats with molecules containing mixed fatty acid residues [10,23-29]. The data for the phase behavior of monoacid TAG mixtures are relatively limited and include phase diagrams for several binary mixtures [18,21,30-36]. We are not aware of structural data in the literature about the phase behavior of multicomponent mixtures of monoacid TAGs.

In the studies about binary TAG mixtures, different types of phase behavior have been reported, depending on the chain-length mismatch of the mixed molecules. For binary mixtures with a chain-length difference of 2 C-atoms, such as LLL+MMM [10,18] and PPP+SSS [21,30,31,34,36], the formation of completely miscible solid solution phases in α and β' polymorphs was reported, whereas a phase separation and eutectic phase were reported for the most stable β polymorphs. In contrast, the phase diagrams for mixtures with bigger chain-length differences are rather complex [10,18,24]. Partial immiscibility was reported even for the α and β' polymorphs in LLL+PPP and LLL+SSS mixtures [18,24]. Further discussion of the properties of the binary TAG mixtures is presented in Section 3.1.2, after presenting our new results.

The major aim of the current study is to investigate the role of the: (1) TAG mixture composition and (2) cooling protocol on the polymorphic phases obtained upon crystallization of monoacid triglyceride mixtures, and (3) related polymorphic phase transitions occurring upon subsequent heating of the solidified samples. Twelve different mixtures were prepared and studied by differential scanning calorimetry (DSC) and small-angle and wide-angle X-ray spectroscopy (SAXS and WAXS). The obtained results allowed us to define several general rules for the phase behavior, and to expand and refine the current understanding of the crystallization and melting processes in such mixtures.

The paper is organized as follows: Section 2 presents the materials and methods used. Section 3 is divided into two main subsections – the phase behavior of various LLL+MMM binary mixtures is discussed in Section 3.1, whereas the main results for the other studied mixtures are presented in Section 3.2. The main conclusions are summarized in Section 4.

## 2. Materials & methods
### 2.1 Materials

Various (in-house prepared) mixtures of TAGs with monoacid chains were studied, see Table 1. They are denoted as "*MXy*", where "*X*" shows the number of triglycerides in the mixture and the letter "*y*" (varied between *a* and *f*) is used when several mixtures with a given number of triglycerides are studied. The TAG chain-length in the mixtures was varied between 8 and 16 C-atoms for the triglycerides with saturated chains, namely from tricaprylin ($C_8TG$, CaCaCa) to tripalmitin ($C_{16}TAG$, PPP). We also studied triolein ($C_{18:1}TAG$, OOO) which has three unsaturated C18 chains with a double-bond in the middle of each chain and the natural coconut oil (CNO, produced under the brand Dragon Superfoods by Smart Organic, Bulgaria). Tricaprylin (purity ≥ 97 %), tripalmitin (≥ 85 %) and triolein (≈ 65%) were purchased from Sigma-Aldrich, while the other triglycerides studied – tricaprin (purity > 98%), trilaurin (>98%) and trimyristin (> 95%) were products of TCI Chemicals. All TAGs were used as received and the studied mixtures were prepared by weighting the necessary amount of each triglyceride. The mixtures were homogenized by melting. The total fatty acid composition of the most complex mixture, *M6*, resembles that of the natural CNO. However, the main fraction of the triglycerides in CNO has mixed chains [37], whereas all triglycerides in the model mixture studied are with monoacid chains, thus allowing us to compare the behavior of monoacid *vs.* mixed-acid TAG mixtures.

**Table 1.** **Composition of the studied TAG mixtures and notation used throughout the text.**

| Triglycerides notation | | Monoacid triglyceride content in the model mixtures, wt. % | | | | | | | | | | | |
|---|---|---|---|---|---|---|---|---|---|---|---|---|---|
| | | *M2a* | *M2b* | *M2c* | *M2d* | *M2e* | *M2f* | *M3a* | *M3b* | *M4* | *M5a* | *M5b* | *M6* |
| $C_8$TAG | CaCaCa | | | | | | | | | | 7 | | 7 |
| $C_{10}$TAG | CCC | | | | | | | | | 15 | 15 | 15 | 8 |
| $C_{12}$TAG | LLL | 50 | 70 | 30 | 85 | 15 | 77.7 | 33.33 | 60 | 50 | 50 | 50 | 50 |
| $C_{14}$TAG | MMM | 50 | 30 | 70 | 15 | 85 | 22.3 | 33.33 | 25 | 20 | 18 | 18 | 18 |
| $C_{16}$TAG | PPP | | | | | | | 33.33 | 15 | 15 | 10 | 10 | 10 |
| $C_{18:1}$TAG | OOO | | | | | | | | | | 7 | | 7 |

### 2.2 Methods

*DSC*

DSC experiments were performed with a Discovery DSC 250 apparatus (TA Instruments, USA). The samples were placed in Tzero aluminum pans, weighted, closed with Tzero aluminum lids and then sealed hermetically using a Tzero sample press. The cooling rate was varied between 1 and 20°C/min, whereas the heating rate was fixed at 1°C/min in most of the experiments. Before each cooling step, the samples were pre-melted at 60 or 70°C and kept at this temperature for 5 min to ensure that the crystal memory of the samples is erased. In preliminary experiments we tested also longer storage periods (up to 1 h) at high temperature, but we did not observe any significant difference between the signals detected after 5 min or 1 h storage in the molten state. Both the cooling and the heating curves were recorded and then analyzed using the build-in functions of the TRIOS data analysis software (TA Instruments). These experiments gave us qualitative information about the phase transitions occurring upon temperature variation in the samples and quantitative information about the enthalpies of each phase transition.

*Small- and wide- angle X-ray scattering (SWAXS)*

To study the molecular order of the triglyceride molecules in details, we performed time and temperature-resolved SAXS and WAXS measurements (denoted below as SWAXS) using Xeuss 3.0 equipment (Xenocs, France) with Cu K-α radiation source, λ ≈ 1.54 Å. The scattered signal was detected with an EIGER2 4M detector (Dectris AG, Switzerland) at a sample-to-detector distance of 286.5 mm. The Xeuss 3.0 equipment allows one to work without a beam stopper included, with the transmitted signal of the direct beam being also measured. The latter is used for intensity calibration of the obtained 1D profiles in the so-called "absolute intensity" mode, which allows a direct comparison of the intensities in the separate experiments. The exposure time for a single image varied between 30 and 300 s, and a 60 s exposure time was used in most experiments. The temperature of the samples was precisely controlled using a HFSX350 high temperature stage, equipped with T96 temperature controller and an LNP96 liquid nitrogen pump, all products of Linkam Scientific Instruments Ltd., UK.

Samples were placed in cylindrical borosilicate glass capillary tubes with outer diameter of 1 mm and 0.01 mm wall thickness, a product of WJM-Glass, Müller GmbH, Germany. Three cooling protocols were applied in most experiments to clarify the effect of the cooling rate on the phases formed upon crystallization: (1) Controlled cooling with 1 or 1.5°C/min rate from the molten sample. This protocol is referred within the text as "*slow cooling*"; No any detectable difference was observed with these two cooling rates. (2) Controlled cooling with 20°C/min from molten sample; and (3) "*Rapid cooling*" protocol which involved pre-cooling of the temperature stage down to -20°C or -50°C and then inserting the pre-melted sample into the cold stage. In this case, the temperature of the sample decreased from $T \approx +70°C$ to the given subzero temperature in a few seconds only, *i.e.* the cooling rate in these experiments was approximately 20-30°C/s (*viz.* 1200 to 1800°C/min). In the text we refer to this cooling protocol as "rapid cooling" or "25°C/s". In all experiments, heating rate of 1.5°C/min was used unless otherwise noted. All measurements were carried out at least in duplicate. All samples were kept in a molten state for at least 5 min before the cooling onset to ensure that the crystal memory of the samples has been erased. In the preliminary series of experiments, we checked also longer periods for storage in a molten state, but we did not detect any difference between these experiments and the final protocol chosen.

The scattering due to the sample holder was subtracted to obtain the correct scattering intensity. To obtain the peak positions, we performed peak deconvolution analysis using Gaussian functions. The scattering profiles presented in the article are shifted with respect to the *y*-axis for clarity. All *d*-spacing values in the text are calculated by averaging the data from the first and the third-order reflections from at least two independent experiments.

## 3. Results and discussion
### 3.1 Two-component LLL+MMM mixtures of various compositions
#### 3.1.1 *Crystallization of M2a mixture upon rapid cooling*

The DSC and SWAXS results for *M2a* mixture (1:1 LLL+MMM) are presented in Figure 2 and in Supporting Figures S1-S3. The DSC signal obtained upon 20°C/min cooling from melt shows a broad peak between 22.7 and -5°C with two maxima at 18.9 and 13.3°C, see the blue dash-dot curve in Figure 2a. These two maxima indicate that two consecutive freezing events occurred in this sample.

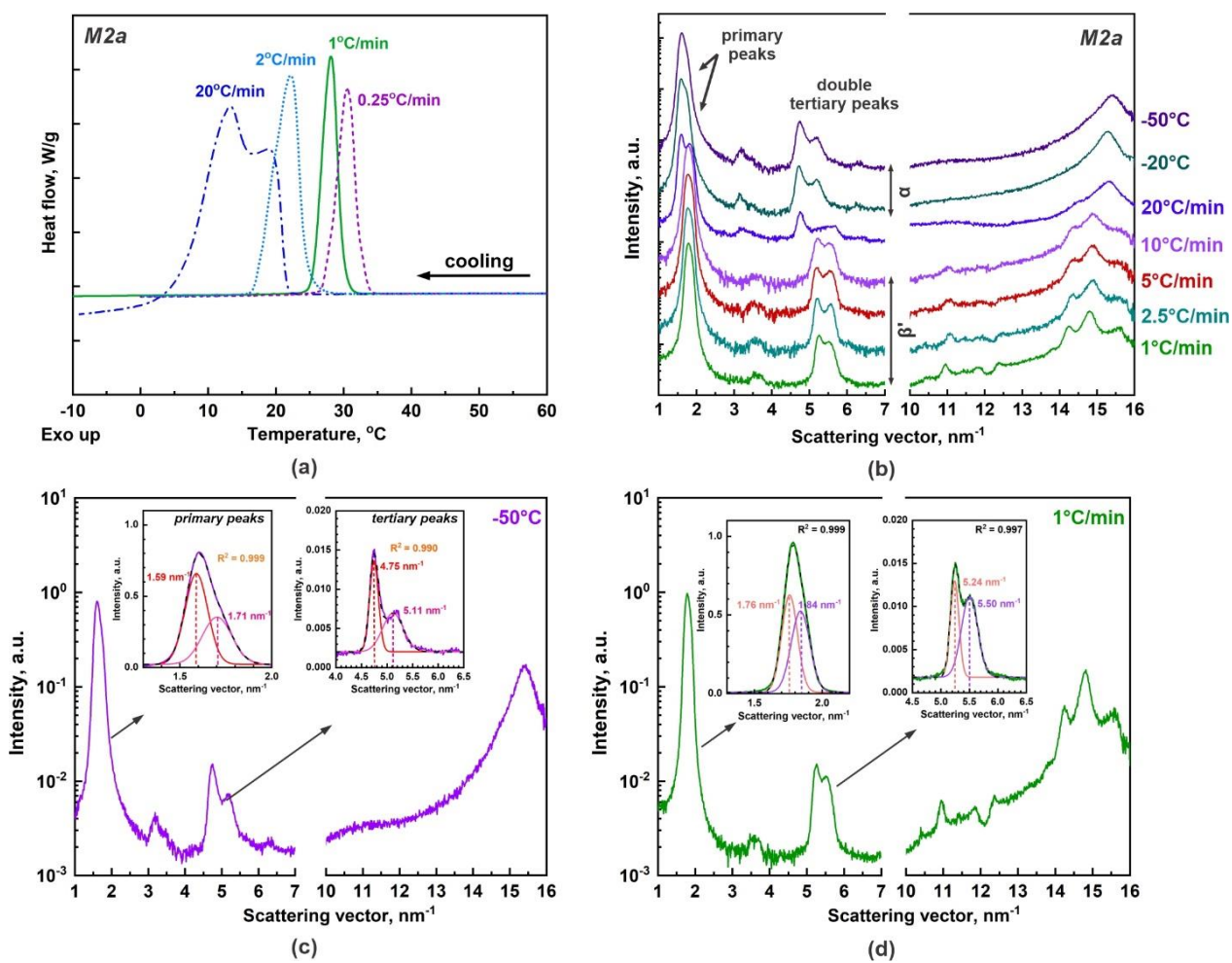

**Figure 2.** **(a)** DSC and **(b-d)** SWAXS profiles for *M2a* mixture of 1:1 LLL+MMM. **(a)** Single exothermic peak is observed upon slow cooling (2°C/min, light blue dashed curve; 1°C/min, green solid curve; 0.25°C/min, purple dashed curve), whereas two crystallization maxima are observed upon cooling at 20°C/min (dot-dashed blue curve). Note that the peak formed at 2°C/min has asymmetric shape. **(b)** Effect of cooling rate on the phases formed upon crystallization: β' phases form when 10°C/min or lower cooling rate is applied, whereas α-phases are observed at higher cooling rates or when a molten sample is inserted into a precooled stage (results for both -20°C and -50°C are shown in the figure). Two mixed coexisting phases are observed in all these experiments. **(c,d)** SWAXS curves obtained upon (c) rapid cooling to -50°C and (d) slow cooling at 1°C/min. Enlarged pictures of the primary and tertiary peaks, along with their deconvolution analysis, are shown as insets. See text for more detailed explanations.

The structural data obtained with the highest cooling rate (rapid cooling down to -20°C or to -50°C) confirmed that two coexisting phases with different lamellar spacings were formed, see Figure 2c. The primary peak, observed at scattering vector $q \approx 1.3$-$2.0$ nm$^{-1}$, had an asymmetric shape due to the partial overlap of the peaks arising from the two distinct phases. Their lamellar spacings can be obtained after performing peak deconvolution analysis, leading to $d_1 = 3.98$ nm ($q \approx 1.58$ nm$^{-1}$) and $d_2 = 3.69$ nm ($q \approx 1.71$ nm$^{-1}$). The presence of these two phases is confirmed also by the tertiary peaks, in which the maxima are better separated: the separation of 0.13 nm$^{-1}$ between the two maxima in the primary peak is tripled in the tertiary peak. At high $q$ values, the spectra show a single broad peak with maximum at 15.3 nm$^{-1}$, characteristic of the hexagonal α phase. Therefore, the two coexisting phases in this sample are two immiscible α polymorphs.

The lamella spacings in these two phases are intermediate between the spacings in the α phases of the pure triglycerides, $d_{\alpha LLL} \approx 3.55$ nm and $d_{\alpha MMM} \approx 4.07$ nm. Hence, both α phases should contain LLL and MMM molecules, mixed in different ratios. The phase with a larger spacing should contain predominantly MMM molecules with a smaller fraction of LLL molecules, whereas the opposite is valid for the phase with a shorter spacing. The fractions of these two phases are approximately equal, as their peak areas are similar. To make the latter conclusion we assume that the specific response factors [13] for the signal obtained from the different domains are similar, because the LLL and MMM molecules belong to the same homologous series.

As previously suggested by Pizzirusso and coauthors [34], the most probable explanation for the intermediate spacing in the mixed domains is that the terminal ethyl groups of the longer fatty acid residues cause a "steric repulsion" between the adjacent lamellae. This steric repulsion increases the distance between the corresponding planes in the mixed phase when compared to the pure phase of the shorter triglyceride molecules, *viz.* from 3.55 for pure $\alpha_{LLL}$ to 3.69 nm for the LLL-enriched phase in the 1:1 LLL+MMM. As expected, the lamellar spacing of 3.98 nm of the MMM-enriched α phase in the same 1:1 LLL+MMM mixture is intermediate between that of the mixed LLL-enriched α phase (3.69 nm) and the pure $\alpha_{MMM}$ phase (4.07 nm). A schematic presentation of these explanations is shown in Figure 3a-c.

3.1.2 *Effect of the cooling rate for the phases formed upon cooling of M2a*

We tested various cooling rates between 1°C/min (0.017°C/s) and *ca.* 25°C/s and always observed crystallization of 1:1 LLL+MMM sample (*M2a*) into two coexisting phases, see Figure 2b-d. The type of the obtained phases depended on the cooling rate. Complete crystallization into the least thermodynamically stable α-phase was observed only with the rapid cooling protocol (25°C/s) which ensures sample crystallization within several seconds. At the intermediate cooling rate of 20°C/min we obtained predominantly α domains. However, at this lower cooling rate (compared to the rapid cooling) we observed also a small fraction of β' domains as the WAXS peak was slightly asymmetric, see Supporting Figure S1b.

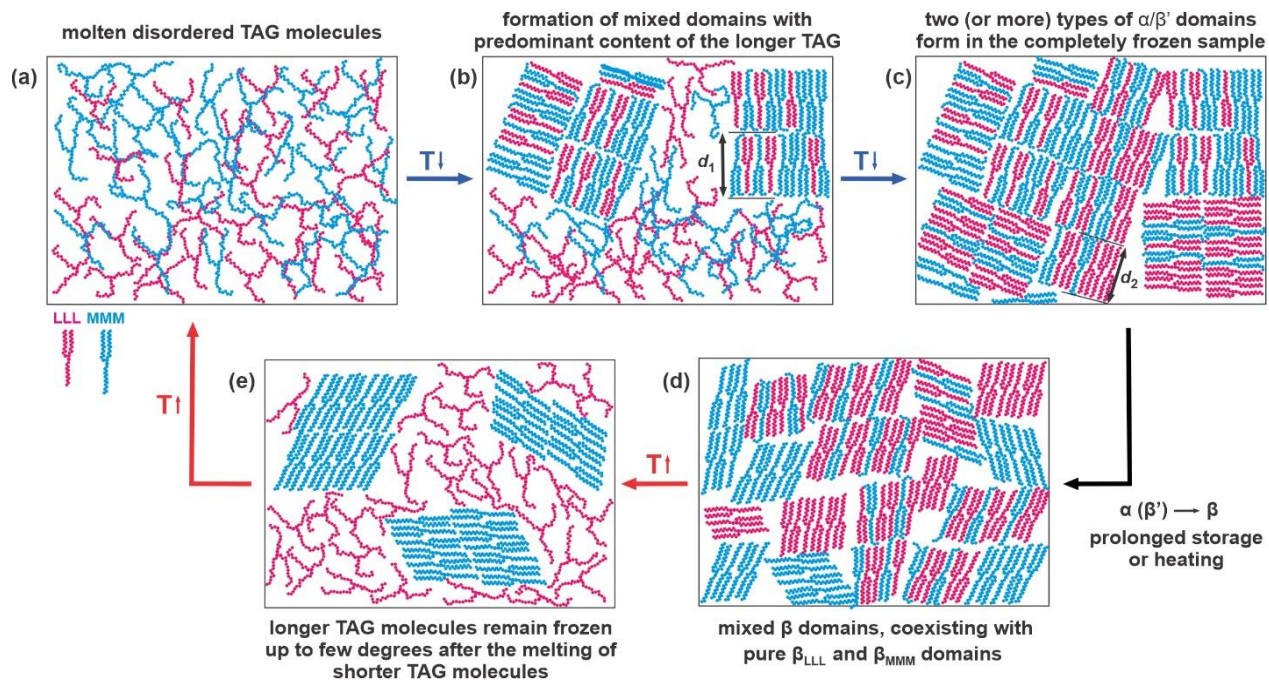

**Figure 3.** **Schematic presentation of the molecular rearrangements observed upon cooling and heating of binary monoacid LLL+MMM mixture.** See text for detailed explanations.

Even lower cooling rates of 1 to 10°C/min resulted in the formation of β' phases with an intermediate stability, see Figures 2d and 4a. The lamellar spacing in these β' phases, 3.57 nm and 3.41 nm, was intermediate between those in the β' phases of pure TAGs, $d_{β'LLL}$ = 3.26 nm and $d_{β'MMM}$ = 3.71 nm. Therefore, these two phases also presented mixtures of LLL and MMM molecules. The presence of these two phases is unambiguously evidenced by the third-order reflection peaks in SAXS spectra, which exhibit two distinct maxima, see the inset graphs in Figure 2d.

The crystallization of these two coexisting β' phases occurred at similar temperatures upon slow cooling, see the SWAXS spectra shown in Figure 4a and Supporting Figure S3a. For that reason, the exothermic DSC peak had a Gaussian shape at a cooling rate ≤ 1°C/min, see Figure 2a. However, the peak became asymmetric at the higher cooling rate of 2°C/min, due to the two consecutive crystallization processes: initially, an MMM-enriched phase with a larger lamellar spacing crystallized, followed by the crystallization of an LLL-enriched phase. This conclusion was confirmed by X-ray scattering experiments at a fixed temperature, see Figure 4c,d. In this experiment we cooled the sample from 70°C down to 33°C at 1°C/min and then observed its crystallization at 33°C for a period longer than 1 h. Initially, a peak centered at a low $q$-vector appeared, evidencing the formation of the phase with a larger spacing, $d \approx 3.57$ nm. After a few minutes, second peak at a higher $q$-vector appeared corresponding to the formation of LLL-enriched phase with $d \approx 3.41$ nm. The primary peak maximum shifted to a slightly higher scattering vector (due to the formation of the second phase) and two distinguishable maxima were formed in the ternary peak, Figure 4c,d. The WAXS spectra evidenced that these two phases were of type β'.

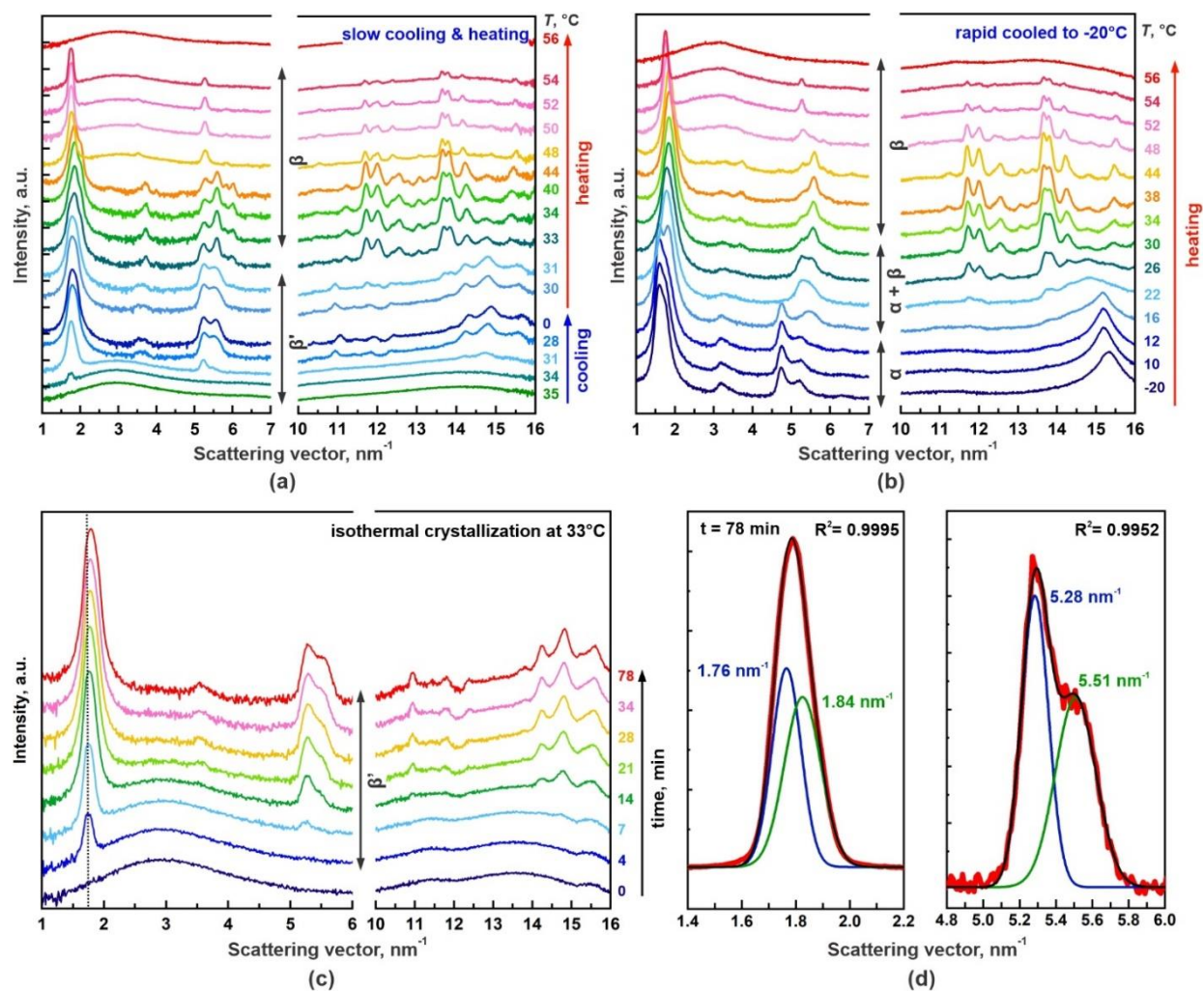

**Figure 4. SWAXS profiles for *M2a* mixture (1:1 LLL+MMM) obtained upon cooling and heating.** (a) SWAXS curves obtained upon cooling and heating at 1°C/min rate. (b) SWAXS curves obtained upon 1.5°C/min heating after rapid cooling to -20°C. The full spectra obtained upon cooling and heating are shown in Supporting Figure S3. (c) Isothermal crystallization of *M2a* mixture at 33°C (this sample had been precooled from 70°C to 33°C at 1°C/min cooling rate). The numbers next to the right-hand axis show the time in minutes, elapsed after the constant temperature has been established. Note that the phase with a predominant content of the longer MMM molecules appears first with peak at $q \approx 1.76$ nm$^{-1}$, whereas longer time is needed for the LLL-enriched phase to crystallize with a shorter lamellar spacing (see the second peak at $q \approx 1.84$ nm$^{-1}$). No changes in the SWAXS patterns are observed between 34 and 78 min (curves not shown). (d) Peak deconvolution analysis for the first and third-order reflections observed after 78 min of isothermal storage at 33°C. All molecules arrange in two distinct β' phases.

One should note that our results (partially) contradict the interpretation given by Takeuchi et al. [18] about the phase diagram of the same LLL+MMM mixture. These authors concluded in Ref. [18] that α and β' phases are miscible, forming solid-solution phases in the entire concentration range, whereas the stable β phases exhibited a eutectic behavior. These conclusions were based on experiments with a protocol including 100°C/min cooling from 100°C to 0°C. This specific cooling rate was not accessible in our experimental equipment. However, as explained above, our experiments with rapid cooling to subzero temperature showed unambiguously that two coexisting α polymorphs were formed in the samples even at very high cooling rates (*ca.* 25°C/s = 1500°C/min), Figure 2c. We note that the spectra for the LLL+MMM system presented in Ref. [18] show only the overlapping primary peaks, whereas the third-order reflection peaks, in which the two coexisting phases can be resolved, appears at scattering angles between 6° and 8.5° which are not shown in Ref. [18]. Therefore, we expect that the experiments described by Takeuchi et al. [18] most probably also involved the formation of two mixed α-phases (LLL-enriched and MMM-enriched), as found in the current study.

The results from coarse-grained molecular dynamics simulations made by Pizzirusso et al. [33] for PPP+SSS mixtures suggest that the molecules remain randomly distributed when α phase is formed. However, these simulations correspond to cooling rates which are by 7 orders of magnitude higher than those achieved in our experimental study. Indeed, the sample is quenched within 450 ns from 173°C to -73 or -98°C in the simulations, which corresponds to a cooling rate of ≈ $3.5 \times 10^{10}$ °C/min. Therefore, the formation of completely miscible α phases could be possible under extremely fast cooling. In our experiments, however, two coexisting α phases were always observed to form.

We note that data for possible formation of two coexisting α polymorphs can be found in the study by Himawan et al. [31] about the phase behavior of PPP+SSS mixtures. However, these authors concluded that the driving force for the observed immiscibility is rather small [31]. No structural data about the spacing or the amount of the two coexisting α phases are presented in Ref. [31].

In summary, the obtained results show unambiguously that two coexisting phases with different molecular compositions and different lamellar spacings form upon crystallization in a 1:1 mixture of LLL+MMM. In the present study, the cooling rate was varied within more than 3 orders of magnitude – from 1°C/min to *ca.* 25°C/s. Cooling rates ≥ 30°C/min were needed to obtain α polymorphs, whereas crystallization into β' phases was observed at lower cooling rates. Below, we discuss how the spacing in these coexisting phases depends on the composition of the LLL+MMM mixture.

3.1.3 *Effect of the LLL+MMM mixture composition on the phases formed upon cooling*

The effect of mixture composition was studied with six mixtures containing LLL and MMM of different ratios: 15:85; 30:70; 50:50; 70:30; 78:22; 85:15. Three cooling protocols were applied – controlled cooling with 20 and 1.5°C/min rates and rapid cooling to -20°C with rate ≈ 25°C/s. These protocols allow the formation of α and β' phases as shown in Sections 3.1.1 and 3.1.2. The mixture composition changed the specific lamellar spacing for these α and β' phases, while the qualitative behavior of the samples remained similar to that described above for the *M2a* mixture (1:1 LLL+MMM).

Figure 5 presents the characteristic lamellar spacings measured in the phases formed in the various mixtures, as a function of LLL fraction in the mixture; see also Supporting Figure S1 for the original SWAXS curves. Only α coexisting phases formed upon rapid cooling, Figure 5a. Their lamellar spacing was intermediate between those in the pure $α_{MMM}$ and $α_{LLL}$ polymorphs. The increase of LLL fraction in the mixture decreased the lamellar spacing of the phases, as expected. The relative fraction of the two phases varied approximately linearly with the change in the mixture composition for LLL ≤ 80 wt.%, whereas the sample with 85 wt.% LLL deviated very significantly from this trend, see Figure 5b.

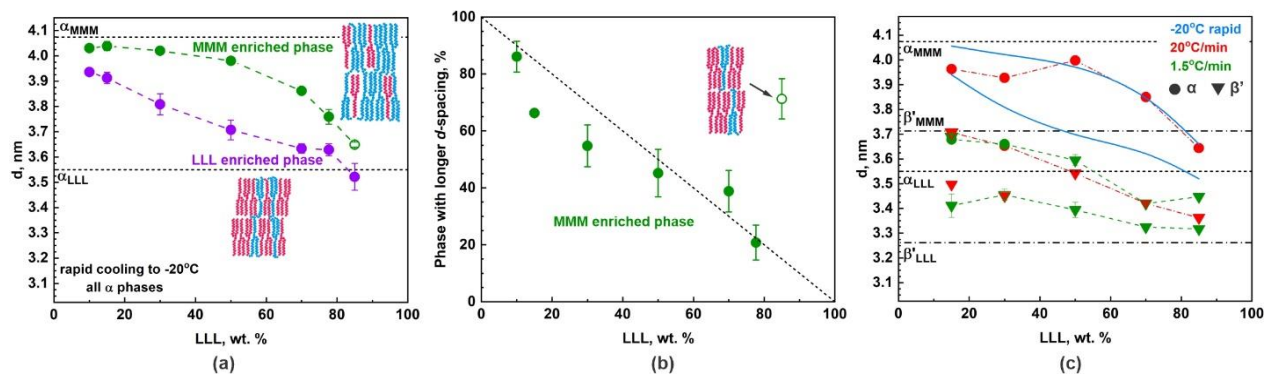

**Figure 5. Lamellar spacing in the various polymorphic phases in LLL+MMM mixtures. (a)** Mixed α coexisting phases formed upon rapid cooling to -20°C. **(b)** Fraction of the phase with longer *d*-spacing formed after rapid cooling. **(c)** Effect of cooling rate on the *d*-spacing: 20°C/min (red symbols) and 1.5°C/min (green symbols). The blue curves are given for comparison and represent the trends shown in (a) for rapid cooling. The circles show phases which are of type α, and the reversed triangles – of type β'. The dashed horizontal lines in all graphs represent the spacings for the respective polymorphs of the pure TAGs. Two or three data points are shown at given LLL fraction and cooling rate – these points represent data for coexisting phases. See the text for more detailed explanations.

Two phases with characteristic spacings of 3.65 ± 0.01 nm and 3.52 ± 0.05 nm were observed in the sample containing 85 wt. % LLL. Their peak areas were 71 ± 7 % and 29 ± 7 %, respectively. The spacing of 3.52 nm indicates that the corresponding phase is pure $\alpha_{LLL}$ phase. It coexisted with mixed LLL+MMM phase which should contain *ca.* 56 wt. % LLL and 15 wt. % MMM present in this mixture. Therefore, the ratio between the LLL and MMM molecules in the mixed domains is ≈ 4:1. In other words, these LLL-rich mixed domains coexisted with domains containing LLL molecules only. This result suggests that there could be some optimal ratio between the LLL and MMM molecules for efficient incorporation of the longer MMM molecules into an α-phase with a predominant LLL content, *viz.* LLL:MMM ≈ 4:1.

Next, we studied the effect of cooling rate on the type of phases formed upon TAGs crystallization. The SWAXS curves obtained upon cooling at 20°C/min showed the formation of α and β' coexisting phases (compare the WAXS curves in Figures S1a and S1b). The formation of β' polymorph was confirmed also by the measured lamellar spacing – phase with a shorter spacing than both $\alpha_{LLL}$ and $\alpha_{MMM}$ was present in these samples. Therefore, this phase cannot be of type α, see the reversed red triangles in Figure 5c. In *M2a* mixture (1:1 LLL+MMM) the phase with a shorter *d* had very similar spacing to that of α phase in pure LLL, see the reversed red triangle in Figure 5c for 50 wt. % LLL. Nevertheless, the WAXS peak was asymmetric in this sample, thus indicating the presence of β' domains as well.

Interestingly, the phase with the largest lamellar spacing obtained upon cooling at 20°C/min was identical to the MMM-enriched phase formed upon rapid cooling (≈ 25°C/s), when the LLL fraction was ≥ 50 wt. % – see the uppermost red circles in Figure 5c which coincide with the upper blue line. In the mixtures with a predominant MMM content, the respective spacing was a bit shorter (by *ca.* 0.18 ± 0.05 nm) compared to that observed in the same samples upon rapid cooling. These results evidence that the reduction of the cooling rate from ca. 25°C/s to 20°C/min affects only the arrangement of the shorter LLL molecules. In contrast, the longer MMM molecules adopted similar configuration at these both cooling rates. This difference in behavior could be explained by considering the different mobility of the two molecular species – the additional time available prior to nucleation and crystal growth at 20°C/min cooling appears to be sufficient for better rearrangement of the shorter LLL molecules and for formation of a more ordered phase. The bigger MMM molecules need longer time for rearrangement and, hence, they acquire similar packing at these two cooling rates.

The interpretation of the three distinct maxima, observed in the third-order reflection peak for the samples with 15 and 30 wt.% LLL after 20°C/min cooling, is more complicated. The spectra for these samples are shown with red and violet curves in Supporting Figure S1b and the respective spacings are shown with red symbols in Figure 5c. The presence of three maxima suggests that (at least) three distinct phases form upon crystallization in these samples. Note that the respective spacings are very similar in these two samples. The phase with largest spacing ($d \approx 3.95$ nm) should be a mixed $\alpha_{LLL/MMM}$ polymorph, because this spacing is larger than the lamellar thicknesses of both β' phases of the pure TAGs, see the red circles in Figure 5c. This phase has a peak area of ≈ 19% from the total peak area. Similarly, the peak observed at the highest *q*-value can be attributed to mixed LLL-MMM β' polymorph, because the corresponding spacing $d \approx 3.47$ nm is shorter than those of the α polymorphs in pure LLL and MMM. The peak area for this phase is 20 ± 5% or 37 ± 11% from the total peak area for the 15 or 30 wt. % LLL mixtures, respectively. The main fraction of TAG molecules in these two samples (15 and 30 wt. % LLL) is included in the phase with a spacing ≈ 3.70 nm (peak area of ≈ 62% for 15 wt. % LLL sample and 53% for 30 wt. % LLL sample). To identify the nature of these peaks we first *assumed* that this is a phase of type α. Then we compared the measured spacings to those of the mixed LLL+MMM α-polymorphs, see the blue curves in Figure 5c. The intersection points between the horizontal line drawn at *d* = 3.70 nm and the blue lines will be at LLL:MMM ≈ 54:46 and LLL:MMM ≈ 83:17 for the two coexisting α phases. In other words, mixed α phase with 3.70 nm spacing could be formed at ratio of LLL:MMM ≈ 54:46 or LLL:MMM ≈ 83:17. In the studied mixtures of 15 wt. % and 30 wt. % LLL molecules, however, the fraction of LLL is much smaller compared to that of MMM. Thus, we conclude that our initial assumption, that the phases with $d \approx 3.70$ nm formed in the 15 wt. % and 30 wt. % LLL mixtures are of type α, contradicts the experimental data.

The next simplest explanation would be that the phases with $d \approx 3.70$ nm are β'$_{MMM}$ phases. However, similar phases were observed also in the experiments with slower cooling rate (1.5°C/min), see the coinciding green and red symbols for 15 and 30 wt. % LLL in Figure 5c. The WAXS curves for the samples cooled at 1.5 and 20°C/min were completely identical to each other and very different from those observed at higher LLL fractions, see Supporting Figure S1b,c. Therefore, even upon slower cooling, a fraction of the molecules in these LLL+MMM mixtures arrange in α-polymorph with a spacing of $d \approx 3.67$ nm, whereas the remaining molecules form pure β'$_{MMM}$ polymorph with very similar spacing. Then, the final conclusion for the peak observed at $q \approx 1.70$ nm$^{-1}$ ($d \approx 3.70$ nm) at 20°C/min cooling of samples with 15 or 30 wt. % LLL is that it is caused by the presence of both α mixed LLL-MMM phase and a pure β'$_{MMM}$ phase.

The same interpretation applies also for the phases formed at slower cooling, 1.5°C/min. In this case the areas of the peaks with $d \approx 3.67$ nm were 88% and 70% for the samples with 15 and 30 wt. % LLL, respectively. Therefore, at least 3 phases (one of type α and two of type β') should be present in these samples to explain the observed scattering signals, despite the fact that two main maxima are observed only, see Supporting Figure S1c.

In the other experiments with slow cooling (1.5°C/min), we observed the formation of β' polymorphs in the samples containing ≥ 50 wt. % LLL, whereas in the samples with a predominant fraction of MMM, we observed coexisting α and β' phases. The phase with a larger *d*-spacing, formed at 1.5°C/min, turned out to be similar to the phase with a shorter spacing formed at 20°C/min (see the coinciding red and green reversed triangles in Figure 5c), whereas the second phase formed at 1.5°C/min had shorter lamellar spacing.

Summarizing, two or more coexisting phases form in all systems studied. Under rapid cooling (≈ 25°C/s) from melt, two coexisting α phases are formed. Lower cooling rate ensures longer time for molecular rearrangement and more ordered β' phases are formed. The slow cooling of 1.5°C/min allows the formation of β' polymorphs in the mixtures containing predominantly shorter and more mobile LLL molecules. In contrast, coexistence of α and β' domains is observed in the mixtures containing predominantly MMM, at the same cooling rate. Further decrease of the cooling rate would be needed to obtain samples ordered entirely in β' polymorphs in MMM-rich systems. We do not observe a direct formation of β polymorphs upon cooling of LLL+MMM mixtures from melt at any of the cooling rates tested ≥ 1°C/min.

3.1.4 *Phase behavior of the binary mixtures upon heating*

Upon cooling, all studied binary LLL+MMM mixtures crystallized in the thermodynamically unstable polymorphic modifications, α and/or β'. Therefore, polymorphic phase transitions took place upon prolonged storage and/or upon heating of these samples to form the most stable β polymorph, see Figure 4a,b.

In the current study, we heated the samples at 1 or 1.5°C/min rate, after the preceding cooling stage. The β'→β phase transition in slowly precooled 1:1 LLL+MMM mixture began at temperatures close to the melting temperature of β'$_{LLL}$ ≈ 35°C, see Figure 4a. Similarly, the polymorphic phase transition for the same sample which was rapidly cooled to crystallize in α phase began at temperature close to the melting temperature of α$_{LLL}$ ≈ 15°C, see Figure 4b. These results show that the phase transition temperatures of the mixed phases are close to those of the pure shorter-chain components, included in the respective binary mixture. Similar conclusion was also made for all other multicomponent TAG mixtures studied (see Section 3.2 below).

In the next paragraphs we first describe the structure of the β phases formed after the polymorphic phase transition and then, using the formulated conclusions, we argue that a direct α→β phase transition most probably occurs in the rapidly precooled samples, without the formation of a long living intermediate β' phase, *viz.* the β' phase could appear only as a short-living transient phase in these samples.

*Structure of the β phases*

The cooling/heating history of the samples did not affect the structure of the β phases formed after the polymorphic transitions. Furthermore, we were able to explain the first and the third-order reflections, observed for the β phases in all binary LLL+MMM mixtures studied, using three types of β phases only – pure $β_{LLL}$ ($d \approx 3.15$ nm), pure $β_{MMM}$ ($d \approx 3.57$ nm) and mixed $β_{LLL/MMM}$ phase ($d \approx 3.39$ nm). In all cases studied, the standard deviation in $d$ is estimated to be *ca.* ± 0.015 nm, see Figure 6a,b.

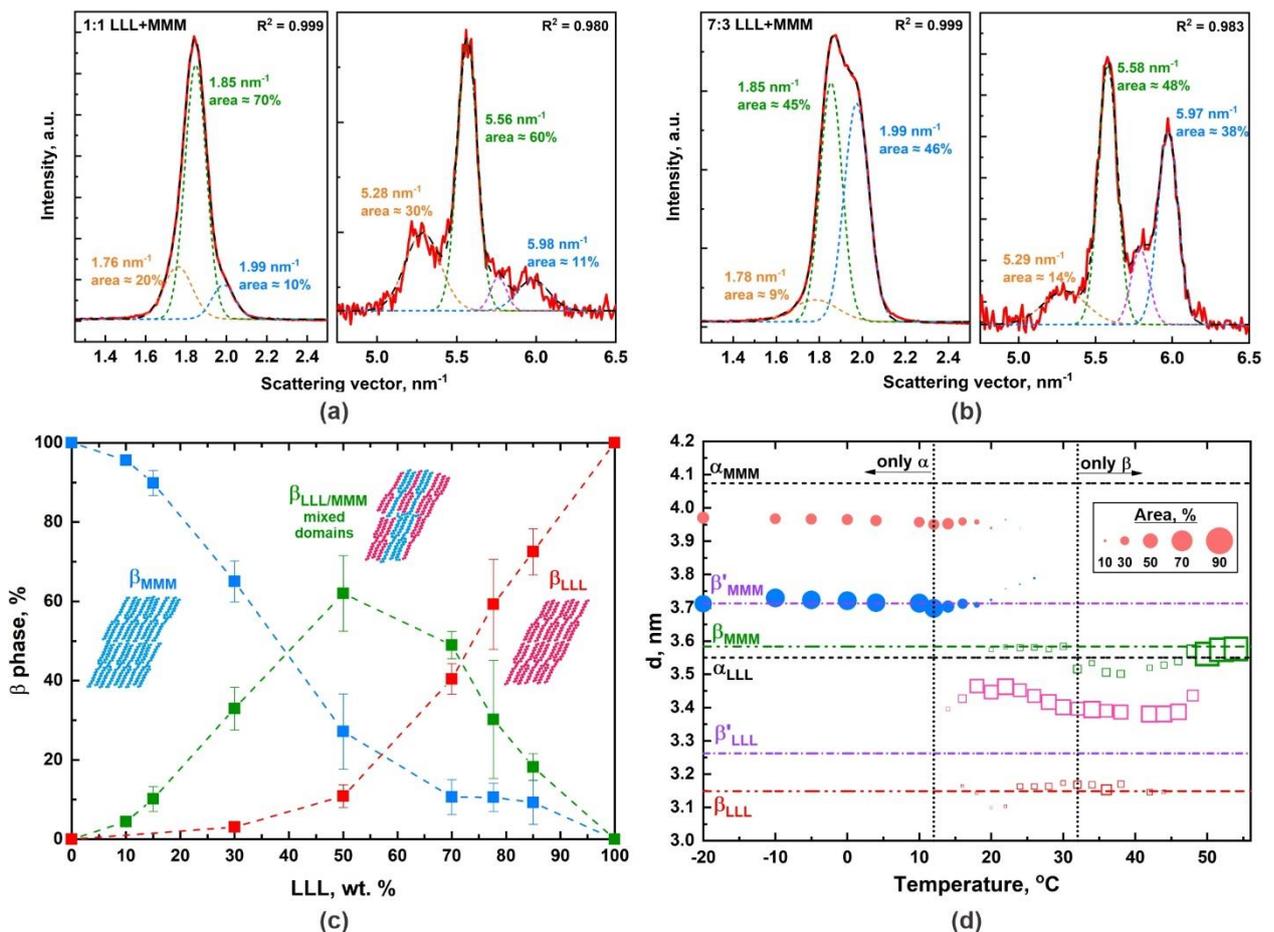

**Figure 6.** **(a,b)** Peak deconvolution analysis for SAXS spectra obtained at 34°C for 1:1 LLL+MMM (a) and 7:3 LLL+MMM samples (b). The peak positions and relative areas (compared to the total peak area) are denoted on the graphs: orange - $β_{MMM}$; green - $β_{LLL/MMM}$ and blue – $β_{LLL}$. The cumulative fit peak is shown with black dashed curves and the red curve represents the original spectra. Note that a fourth peak (shown with purple) is used in the analysis of third-order reflections. This is a mixed reflection peak. Its area has been disregarded in the area analysis. **(c)** Relative content (based on the peak areas) of the different types of β domains, observed in LLL+MMM mixtures. The lamellar spacings are $d \approx 3.15$ nm for pure $β_{LLL}$, $d \approx 3.57$ nm for pure $β_{MMM}$ and $d \approx 3.39$ nm for the mixed $β_{LLL/MMM}$ domains. **(d)** Lamellar spacings as functions of the temperature for rapidly crystallized 1:1 LLL+MMM sample (SWAXS curves are presented in Figure 4b). The circles show α phases and squares – β phases. The size of the symbols denotes the relative area of the respective peek, see the legend inset. The size of the empty red squares has been increased 4 times to make these symbols visible on the graph.

The presence of different β phases was further confirmed by the observations made when the samples were heated above the LLL melting temperature ($T \approx 47°C$). When the melting temperature of the pure $β_{LLL}$ phase was reached, both the small $β_{LLL}$ peak and the peak for the mixed $β_{LLL/MMM}$ phase disappeared, whereas the $β_{MMM}$ peak remained and even increased its absolute area. This observation shows that at least part of the MMM molecules included in the mixed LLL/MMM domains remains in a crystalline state even after the mixed domains disappear in the sample. The $β_{MMM}$ peak remained present in the spectra until the temperature was further increased up to the melting temperature of pure MMM, see Figure 4a,b. Two consecutive melting events were observed also in the DSC thermograms as a wide peak in the range between *ca.* 40 and 57°C, with two well-pronounced maxima showing the melting of the LLL and MMM molecules, respectively, see Supporting Figure S2. These transitions are schematically shown in Figure 3d,e.

The relative fractions of these three phases depended on the mixture composition, see Figure 6c. The mixed $β_{LLL/MMM}$ phase had a maximum in the 1:1 LLL+MMM mixture, while the fractions of the pure $β_{LLL}$ and $β_{MMM}$ phases increased almost linearly with the concentration of the respective components when it became > 50 wt. % in the overall mixture. Small $β_{MMM}$ peaks were observed in all mixtures with a predominant content of LLL. In contrast, pure $β_{LLL}$ peaks were not seen in the samples containing 15 wt. % or less LLL. The latter result indicates that all LLL molecules are included in the mixed $β_{LLL/MMM}$ phase at a much low LLL concentration. Based on the obtained peak areas, one can estimate the ratio of the LLL and MMM molecules in the mixed β domains. For the mixtures with *ca.* 50-80 wt. % LLL, this ratio is LLL:MMM ≈ 2:1, whereas this ratio became much higher for the mixtures with LLL fraction > 80 wt.%. The fact that the *d*-spacing does not change even when the ratio varies from 2:1 to *ca.* 10:1 (in 3:7 LLL:MMM mixture) shows that this spacing is governed mostly by the longer MMM molecules in the mixed $β_{LLL/MMM}$ domains. We note that an overall decrease in the total peak area, as well as disappearance of the mixed $β_{LLL/MMM}$ phase is observed for these mixtures once the LLL melting temperature is exceeded. The average domain size estimated by the Scherrer equation [38] was ≈ 34 ± 16 nm for the $\beta_{LLL}$ domains and ≈ 52 ± 16 nm for both the $\beta_{LLL/MMM}$ and $\beta_{MMM}$ domains.

*α→β transition for rapidly crystallized mixtures*

The analysis of the peak positions and areas for the SWAXS data obtained upon heating for the rapidly crystallized 1:1 LLL+MMM mixture (shown in Figure 4b) is presented in Figure 6d. This sample crystallized into two coexisting α phases – one enriched in LLL and another enriched in MMM molecules. As this is the least stable TAG polymorph, there are two possible consecutive phase transitions which may occur upon heating: α→β' and β'→β. Complete transformation of the α domains into β' domains was certainly not observed in our experiments; instead, WAXS peaks for the β polymorphs appeared at $T ≈ 18\text{-}20°C$, see Supporting Figure S3c. Thus, we tried to analyze the whole transformation sequence assuming the possible presence of 5 different phases in total for the whole temperature range: two α phases and three β phases, as explained above. The analysis shows that such an interpretation is possible if one assumes that the phase transition for a given phase occurs in a given time interval, instead of being instantaneous when a certain temperature is reached. Indications for such a behavior can be seen from the areas of the red and blue circles in Figure 6d. Once $α_{LLL}$ melting temperature is approached, the $α_{LLL/MMM}$ phase (with a higher LLL content) starts to transform into β phase, but the peak for $α_{LLL/MMM}$ does not disappears completely until further heating. Similar results were observed for the other $α_{MMM/LLL}$ phase. We note that some transient short-living β' phase may be present in this sample, but its existence cannot be confirmed by the obtained experimental results, because some of its characteristic peaks in the WAXS range overlap with those for the α and β polymorphs.

In summary, three different β phases were observed in all LLL+MMM binary mixtures after α-to-β or β'-to-β phase transition: pure $β_{LLL}$ phase, mixed $β_{LLL/MMM}$ phase and pure $β_{MMM}$ phase. The relative fractions of these phases depended on the LLL/MMM ratio in the mixture: for LLL ≤ 30 wt. %, the $β_{MMM}$ domains prevailed; for 50 wt. % ≤ LLL ≤ 70 wt. %, the mixed $β_{LLL/MMM}$ phase prevailed; the $β_{LLL}$ phase was the most abundant at LLL > 70 wt. %. A direct α→β phase transition was observed in the samples pre-crystallized in α phases, without the formation of long-living intermediate β' phases.

The phase transitions between the various polymorphs occurred around the melting temperature of the prevailing component in the respective phase, *e.g.* the α→β phase transition started at lower temperatures compared to the respective β'→β transition, as the melting temperature of $\alpha_{LLL}$ phase is lower than that of $\beta'_{LLL}$. The melting of the β phases began with the melting of the $\beta_{LLL}$ and the mixed $\beta_{LLL/MMM}$ phases, after which the amount of the $\beta_{MMM}$ domains increased. The latter coexisted with the already melted shorter LLL molecules within a range of several degrees, until the sample was heated up to the melting temperature of the pure MMM, see schematics in Figure 3.

### 3.2 Phase behavior of complex monoacid triglyceride mixtures

This section presents the results obtained upon cooling and heating of several monoacid TAG mixtures, containing three or more molecular species. The exact composition of these mixtures and their nomenclature are presented in Table 1 above. These mixtures were designed in the following way: we started with the ternary mixture which contained the three main fatty acid residues present in the coconut oil, mixed in equal fractions (1:1:1, mixture *M3a*) and in a ratio similar to that in CNO, *viz.* LLL:MMM:PPP = 60:25:15 (mixture *M3b*). Then we sequentially increased the sample complexity by adding other monoacid TAGs to finally obtain mixture *M6* (CaCaCa:CCC:LLL:MMM:PPP:OOO = 7:8:50:18:10:7) with total fatty acid content equivalent to that of the natural coconut oil (CNO). Thus we studied the effect of triglyceride mixing and compared the obtained structural information to that for CNO. Note, however, that CNO has equivalent total fatty acid content to *M6*, but includes TAG molecules with mixed fatty acid residues. Thus, we could compare the monoacid TAGs with the mixed-acid TAGs of the same fatty acid chemical composition, while having different distributions of the fatty acid residues in the TAG molecules. To better understand the observed trends, several intermediate mixtures were studied as well.

The results for the phase transitions occurring in these samples, cooled at 1.5 and 20°C/min rates, are shown in Figures 7 and 8, Tables 2-3 and Supporting Figures S4-S15. For some of these samples we studied also the phases formed upon rapid cooling to subzero temperatures. Polymorphism was observed in all studied samples. The respective phase transitions were detected as exothermic peaks in the DSC curves (see the green and blue regions in Figure 7b), while the peaks in the scattering experiments changed their positions to reflect the changes in the phases formed.

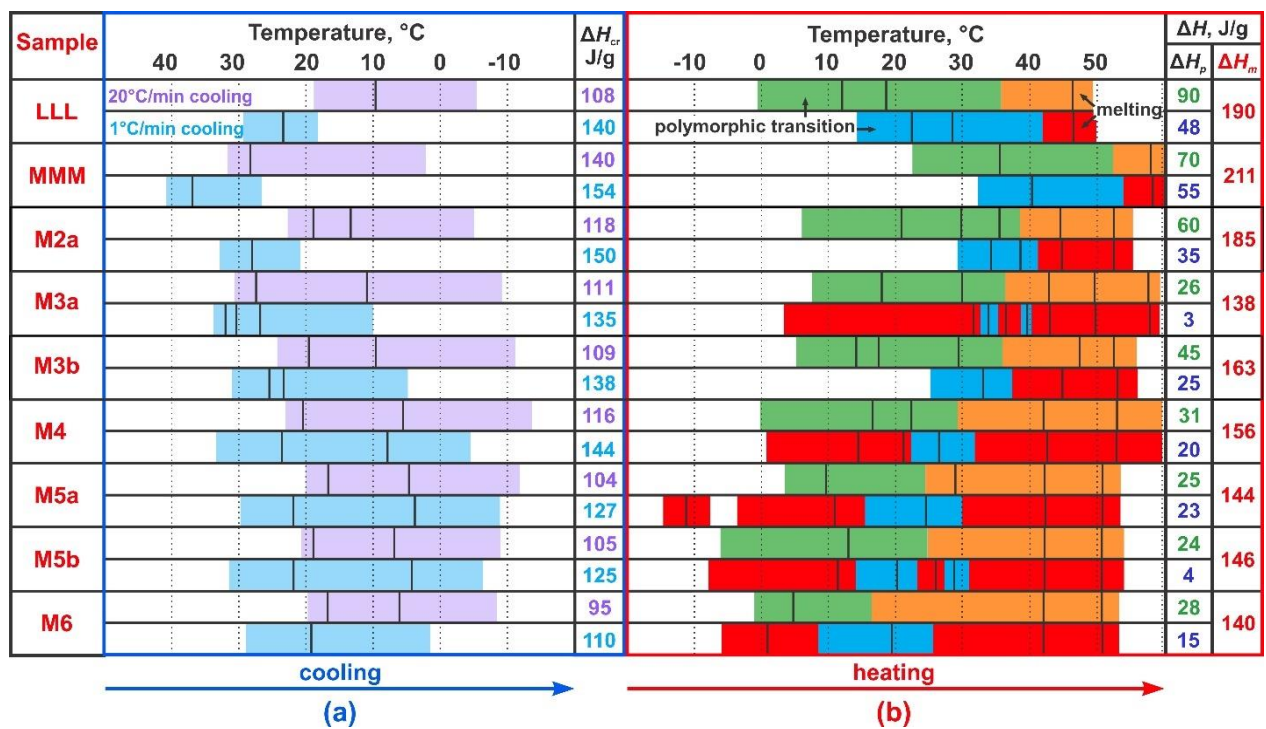

**Figure 7.** **Summary of the phase transitions observed in the DSC experiments with different triglyceride mixtures.** (a) Data, obtained upon cooling. All peaks are exothermic and show freezing processes. (b) Data obtained upon subsequent heating at 1°C/min rate. The green and blue regions represent exothermic peaks, corresponding to polymorphic phase transitions, whereas the orange and red regions represent endothermic peaks, *i.e.* melting processes. (a,b) For each sample, the first row shows the information obtained upon 20°C/min cooling and subsequent 1°C/min heating. The second row shows results obtained upon 1°C/min cooling and heating. The peak maxima are shown with black lines. The enthalpies are denoted with Δ*H* and the subscripts "*cr*", "*p*" and "*m*" correspond to crystallization, polymorphic transition and melting, respectively. The melting enthalpy is average of the two types of experiment with different heating rates and the experimental accuracy is ± 2 J/g. The precise composition of the studied TAG mixtures is presented in Table 1.

3.2.1 *Phases formed upon cooling of multicomponent mixtures*

*Ternary mixtures*

We investigated the phase behavior of two ternary mixtures, composed of LLL, MMM and PPP molecules and mixed in different ratios: LLL:MMM:PPP = 1:1:1 (denoted as *M3a*) and LLL:MMM:PPP = 60:25:15 (*M3b*). Three distinct phases were formed in *M3a* mixture upon rapid cooling to -20°C with characteristic lamella spacings of 3.68, 4.18 and 4.43 nm, see Supporting Figure S4a,b. Their peak areas were approximately 27, 39 and 34% from the total peak area, respectively, *i.e.* the phase with an intermediate thickness was the most abundant one. The WAXS profile showed that all these phases were of type α. A comparison between the measured spacings in the mixed sample and the characteristic spacings for the α polymorphs of pure LLL ($d_{\alpha,LLL} \approx 3.55$ nm), MMM ($d_{\alpha,MMM} \approx 4.07$ nm) and PPP ($d_{\alpha,PPP} \approx 4.54$ nm) shows that all α domains formed upon crystallization of *M3a* mixture should contain mixed TAG molecules. The phase with the largest spacing most probably contains mixed MMM and PPP molecules, whereas the one with the shortest spacing contains a mixture of LLL and MMM molecules. Note that the latter phase is identical to the one formed upon rapid crystallization of LLL:MMM = 50:50 mixture, see the blue circles in Figure 6d above. Analyzing the TAG species distribution using the peak areas, one sees that the mixed phase with an intermediate thickness ($d = 4.18$ nm) should contain all three TAG species present in this mixture, with a predominant fraction of MMM. The PPP should be present because 4.18 nm $> d_{\alpha,MMM} \approx 4.07$ nm, whereas some LLL molecules should be also present as the peak area for the mixed LLL+MMM peak is too small to accounts for all LLL molecules present in this sample.

The SWAXS spectra obtained upon cooling of *M3a* mixture at 20°C/min is presented in Supporting Information Figure S4c. The third-order peak analysis showed the formation of four phases – three phases with a spacings identical to those seen after the rapid cooling ($d = 3.67$ nm, 4.20 nm and 4.42 nm) and an additional phase with a shorter spacing of 3.38 nm.

The latter phase with the spacing of 3.38 nm was a very small fraction with a relative peak area of $\approx 10\%$ only. Although the WAXS signal appeared as characteristic for α polymorph with a single wide symmetric peak centered around $\approx 15.3$ nm$^{-1}$, this phase could not be of type α, as it has a characteristic spacing which is shorter than that for the shortest TAG in the mixture, $d_{\alpha,LLL} \approx 3.55$ nm. Therefore, we conclude that upon 20°C/min cooling, three α phases form in the *M3a* mixture, which coexist with a limited amount of β' phase containing predominantly LLL molecules, mixed with a small fraction of MMM molecules. Note that LLL and MMM have shorter fatty acid residues compared to PPP, thus they rearrange faster upon cooling and, due to their lower melting temperatures, LLL and MMM have longer time available for rearrangement.

*M3a* sample (LLL:MMM:PPP = 1:1:1) crystallized into three coexisting β' phases when cooled at lower rate of 1.5°C/min, see Supporting Figure S4e. These phases had lamellar spacings of 3.35 nm, 3.76 nm and 3.98 nm, which are intermediate between those for the pure β' phases, *i.e.* these phases contain mixed TAG molecular species.

We also investigated the phase behavior of the ternary mixture LLL:MMM:PPP = 60:25:15 (*M3b*). This sample crystallized in four distinct types of α domains when was rapidly cooled, $d = 3.50, 3.63, 3.90$ and 4.23 nm, see Supporting Figure S5a,b. In this case, the 3.50 nm domains were about 10% and contained LLL molecules only, whereas the other three types of domains contained mixtures of molecules. We note that we were able to obtain a good fit to the cumulative peaks using only three distinct peaks. However, in the latter case, the peak areas (45.5; 13.5 and 41 %) did not match with the lamellar spacings (4.24 nm, 4.10 nm and 3.65 nm, respectively) and the TAG composition of the samples. The 4.24 nm phase should contain mainly MMM and PPP molecules, but their total content is 40% in the mixture. At the same time, the phase with the shortest spacing should also contain MMM molecules, because its spacing is larger compared to the pure $\alpha_{LLL}$ phase. Thus, the interpretation with the four distinct types of domains is more realistic.

Four distinct phases were needed to explain the SWAXS curves obtained upon 20°C/min cooling of *M3b* sample with $d \approx$ 4.31, 3.99, 3.57 and 3.35 nm (first order peak maxima at $q \approx$ 1.46, 1.58, 1.77 and 1.87 nm$^{-1}$, respectively), see Figure 8a and Supporting Figure S5a,c. In contrast to *M3a* sample, in this case α domains were with slightly different spacings when compared to those observed after rapid cooling, however, again β' phase was formed ($d \approx$ 3.35 nm). Decrease of the cooling rate to 1.5°C/min led to the formation of three β' phases with $d \approx$ 3.69, 3.41 and 3.30 nm, see Supporting Figure S6. The peak areas were ≈ 4:3:3 for the three peak respectively.

We expect that the molecular mechanism determining the intermediate spacings in mixed TAG domains is similar to the one proposed by Pizzirusso et al. [34] for the binary PPP+SSS mixtures, see the schematics shown in Figure 3. Briefly, a "steric repulsion" is caused by the longer fatty acid residues included in the domains with a predominant content of shorter TAG molecules, thus increasing the overall interlamellar thickness. In the opposite case, when shorter molecules are included in domains with a predominant content of longer TAG molecules, they act as "defects" in the crystalline order of mixed domains and leave free space for inclusion of the terminal groups of neighboring longer molecules, which leads to overall decreased lamellar thickness, see also the related discussion for the redistribution of the molecular species upon cooling and heating presented in Section 3.1.4 above. The only significant difference observed with the more complex mixtures compared to the binary TAG mixtures seems to be the higher number of combinations in which the individual species may mix together.

*Mixtures containing ≥ 4 mixed TAG*

Four monoacid TAG samples containing 4 or more mixed species were studied to verify that our conclusions are not limited to the simpler 2 or 3-component mixtures, but can be applied also to more complex systems. Sample *M4* contained four TAG species: CCC:LLL:MMM:PPP = 15:50:20:15. Upon rapid cooling to subzero temperature, this sample crystallized into four coexisting α phases: $d \approx$ 3.18 nm (10%), 3.62 nm (49%), 4.05 (17%) and 4.33 nm (24%), where the percentage denotes the relative area of the respective peak, see Supporting Figure S8a. The most abundant type of domains ($d \approx 3.62$ nm) was identical to that observed with the 3-component mixtures, see above. Furthermore, the same phase was identified in more complex 5- and 6-component mixtures, see Table 2 and the related discussion below. The main fraction of the shortest CCC molecules was included in the domains with a 3.18 nm spacing.

Four distinct phases were observed also when *M4* sample was cooled at 20°C/min rate, see Supporting Figure S7b. In this case, along with the α domains ($d \approx 4.05$ and 4.38 nm) containing predominantly the longer TAG molecules in the mixtures, β' domains were also formed ($d \approx 2.97$ and 3.32 nm). The most abundant phase was the one with $d \approx 3.32$ nm (peak area $\approx 46\%$). Finally, we tested also 1.5°C/min cooling rate to allow sufficient time for the molecular rearrangement of all different TAG species. Five distinct phases were observed when this protocol was applied: $d \approx 2.91$ nm (8%), 3.29 nm (31%), 3.40 nm (11%), 3.60 nm (32%) and 3.88 nm (18%), see Supporting Figure S7a.

To check whether the long-chain molecules crystallize before the shorter TAGs (similarly to the process observed with 1:1 LLL+MMM, Figure 4c above), we performed isothermal crystallization experiment with this sample (CCC:LLL:MMM:PPP = 15:50:20:15). The molten sample was cooled at 1.5°C/min rate from 70 to 35°C and then the temperature was kept at 35°C for a period of 1 h. As seen from the results shown in Supporting Figure S7c, the crystallization began about 10 min after the constant temperature was established. Initially, a peak with maximum at $q \approx 1.62$ nm$^{-1}$ appeared, *viz.* the phase with the largest *d*-spacing formed first. The peak gradually increased its area, but no further peaks appeared at this temperature for 1 h. The main fraction of TAG molecules remained in a liquid state. Three new peaks appeared in the SAXS spectra when the temperature was decreased down to 27°C, due to the crystallization of the main fraction of LLL and MMM molecules. The CCC molecules remained in a liquid state even after a 30 min storage at this temperature. The peak corresponding to the smallest spacing ($d \approx 2.91$ nm) appeared when the temperature was further decreased down to *ca.* 10°C. This result shows unambiguously that the crystallization behavior of the complex TAG mixtures is qualitatively similar to that observed in the simpler mixtures. However, larger number of phases is formed in such samples, due to the larger number of components mixed.

Very similar phases were formed upon rapid cooling of *M5a*, *M5b* and *M6* samples, see Supporting Figure S8. The main TAGs included in these samples are LLL (50 wt. %), MMM (18 wt. %), PPP (10 wt. %) and CCC (15% in *M5a* and *M5b* and 8 wt. % in *M6*). *M5a* sample contains also 7 wt. % CaCaCa, whereas the latter is exchanged by OOO in *M5b*. *M6* sample contains both 7 wt. % CaCaCa and 7 wt. % OOO. The α phases had slightly shorter spacings compared to the phases formed in *M4* sample, due to the inclusion of shorter-chain molecules, *d*

≈ 3.00 nm (≈ 4% for *M6* and ≈ 6.5% in *M5*), 3.33 nm (≈ 8%), 3.61 nm (≈ 43%), 3.96 (≈ 24.5%) and 4.23 nm (≈ 18%). Most abundant was the phase with 3.61 nm spacing.

Illustrative results about the phases formed upon cooling in *M5a*, *M5b* and *M6* samples upon 1.5°C/min and 20°C/min cooling are presented in Figures 8b,c and Supporting Figures S9-11. Coexisting domains of α and β' type were observed upon 20°C/min cooling, whereas at 1.5°C/min rate β' and β domains were formed. The respective spacings are presented in Table 2. In such complex mixtures, the specific spacing of the β' phases cannot be determined precisely, because no experimental conditions were found to arrange all molecules in β' domains. In the slowly cooled samples, the β' phase coexisted with β domains, whereas in the mixtures cooled at 20°C/min rate α domains always coexisted with α, β and/or liquid phases. We note that although some spectra seemed to have only two main maxima in the first order peaks (see for example Supporting Figure S10a or S11c), the analysis of both the first and third-order reflection peaks revealed four to eight different types of coexisting domains formed upon crystallization. As for the simpler mixtures, these phases had intermediate spacings when compared to those of the pure monoacid TAGs included in the mixtures. As expected, the crystallization always started with the longest TAG species, whereas the shorter ones remained in a fluid state down to much lower temperatures. This is easily seen from the temperature-resolved SAXS curves obtained at 1.5°C/min cooling, see Supporting Figures S9, S10a and S11. In all these spectra, initially the first-order SAXS peaks at lower *q*-vectors appear, whereas additional peaks showing the crystallization of the shorter TAGs appear at lower temperatures.

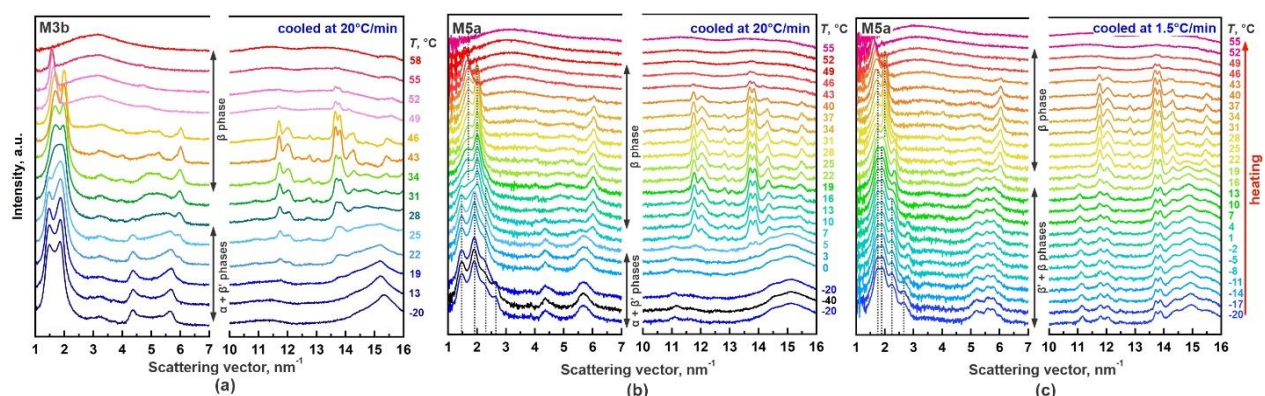

**Figure 8.** **Scattering curves obtained upon heating of model triglyceride mixtures.** **(a)** 1.5°C/min heating of *M3b* sample after 20°C/min cooling. **(b,c)** 1.5°C/min heating of *M5a* sample after (b) 20°C/min and (c) slow 1.5°C/min cooling. Note that the formed phases depend on the cooling protocol. The cooling curves for (c) are shown in Supporting Figure S9. See the main text for more detailed explanations of the scattering spectra. The mixture compositions are shown in Table 1.

The present study is focused on the phase behavior of monoacid TAG mixtures. However, the natural oils and fats contain predominantly TAG molecules with mixed fatty acid residues. To check how our results compare with the respective properties of the natural CNO, we performed SWAXS experiment with CNO as well. We note that the total composition of *M6* mixture resembles the total fatty acid content in CNO. However, our mixtures contain monoacid TAGs only, whereas ≈ 80% of the TAG molecules in CNO have mixed chains [37].

Supporting Figure S12 shows the spectra for CNO. Two main peaks are observed upon rapid cooling from melt with $d \approx 3.30$ and 3.82 nm. The primary peak in *M6* sample could be fitted with these two spacings if the two separate maxima at $q \approx 1.53$ and $1.71$ nm$^{-1}$ are combined into a single maximum at $q \approx 1.65$ nm$^{-1}$. This comparison shows that the averaged lamellar spacing of the phases formed in *M6* sample is similar to that of the phases formed in CNO. However, due to the different fatty acid distribution in the model mixture, the monoacid TAG molecules prefer to pack with each other in separate domains instead of forming a single mixed phase with a larger number of mixed species. Furthermore, the temperature at which the 3.35 nm phase in *M6* undergo polymorphic phase transition, $T \approx 4$-$5°C$, coincides with the temperature for $\alpha + \beta_1' \rightarrow \beta_2'$ phase transition in CNO. However, as the TAG molecules in *M6* sample have monoacid chains, they pack more tightly in β phase which melts at much higher temperature, whereas the mixed-acid chains in CNO allow formation of a single $\beta_2'$ phase for all CNO molecules which then melts at a much lower temperature compared to *M6*, see Table 3.

In contrast, the phases formed upon slow cooling in *M6* mixture with characteristic spacings of 4.12 nm (5%), 3.70 nm (25%), 3.34 nm (25%), 3.15 nm (41%), 2.70 nm (2%) and 2.24 nm (2%) differed significantly from the single thermodynamically stable $\beta_2'$ phase formed in CNO ($d \approx 3.39$ nm). This comparison shows that the monoacid triglyceride molecules can pack much better and phase separate from each other upon slow cooling which is not the case for CNO.

### 3.2.2 *Polymorphic phase transitions in multicomponent mixtures upon heating*

After the initial crystallization of the model mixtures upon cooling, further polymorphic phase transitions occurred in these samples. These transitions began at significantly lower temperatures for the samples which were cooled at a faster rate, because they contained predominantly the least stable α phase. This temperature was *ca.* $16 \pm 3°C$ for 2- and 3-component mixtures, whereas it was ≈ $0 \pm 3°C$ for 5- and 6-component mixtures, see the green regions in Figure 7b. For the samples cooled at lower rates, the temperature at which the polymorphic phase transitions began (in this case β' → β) was significantly higher – compare the green and blue regions in Figure 7b.

In agreement with the obtained structural data, the freezing enthalpies were significantly higher for the slowly cooled samples which arranged in β' or in β'+β phases, compared to the rapidly cooled samples which froze in α or α+β' phases. This difference was compensated when the polymorphic phase transitions occurred to include the enthalpy for the α→β transition. For that reason, the polymorphic transition enthalpy for the quickly cooled mixtures, $\Delta H_p$, was by 10-25 J/g higher than $\Delta H_p$ measured upon heating of slowly precooled samples, see Figure 7b.

Figure 9 presents illustrative SAXS spectra for the β phases formed in the multicomponent mixtures. The structure of these phases did not depend on the thermal history of the samples, as explained above for the binary mixtures – the structure was exclusively determined by the molecular content of the mixtures. As explained in Section 3.1, the LLL+MMM mixtures were found to crystallize always in three coexisting β phases: $β_{LLL}$, $β_{LLL/MMM}$ and $β_{MMM}$ (see section 3.1.4 above). Hence, it was interesting to see whether this conclusion applies also when the mixtures contain more than two molecular species.

**Table 2.** **Structure of β phases formed in TAG mixtures after cooling and subsequent slow heating.** The fraction of the various phases was determined from the β phase spectra obtained after rapid cooling to $T = -20°C$. The average standard deviation of the $d$-values is *ca.* ± 0.01 nm for the pure β phases and ± 0.02 to 0.05 nm for the mixed β phases. The interlamellar spacings are described ($R^2 = 0.999$) by the equation $d_β \approx 0.222\bar{n} + 0.4682$ nm, where $\bar{n}$ is the average number of C-atoms in the fatty acid residues.

|  | $q$, nm$^{-1}$ | $d$, nm | Phase abundance in TAG mixtures, % | | | | | | |
|---|---|---|---|---|---|---|---|---|---|
|  |  |  | *M2a* | *M3a* | *M3b* | *M4* | *M5a* | *M5b* | *M6* |
| $β_{CaCaCa}$ | 2.81 | 2.23 |  |  |  |  | 3 ± 1 |  | 1 ± 0.5 |
| $β_{CaCaCa/CCC}$ | 2.52 | 2.49 |  |  |  | 1 ± 0.5 |  |  | 2 ± 1 |
| $β_{CCC}$ | 2.34 | 2.68 |  |  |  | 2 ± 0.2 | 7 ± 3 | 4 ± 2 | 4 ± 3 |
| $β_{CCC/LLL}$ | 2.19 | 2.88 |  |  |  | 13 ± 7 | 1 ± 0.5 | 2 ± 1 | 3 ± 2 |
| $β_{LLL}$ | 2.00 | 3.15 | 11 ± 3 | 8 ± 2 | 39 ± 4 | 31 ± 8 | 46 ± 3 | 54 ± 5 | 52 ± 2 |
| $β_{LLL/MMM}$ | 1.85 | 3.39 | 62 ± 10 | 13 ± 4 | 12 ± 6 | 12 ± 7 | 5 ± 1 | 6 ± 4 | 9 ± 1 |
| $β_{MMM}$ | 1.75 | 3.57 | 27 ± 10 | 24 ± 9 | 21 ± 2 | 13 ± 4 | 14 ± 12 | 15 ± 1 | 14 ± 1 |
| $β_{MMM/PPP}$ | 1.66 | 3.78 |  | 33 ± 13 | 19 ± 7 | 19 ± 5 | 15 ± 6 | 15 ± 2 | 11 ± 3 |
| $β_{PPP}$ | 1.56 | 4.02 |  | 22 ± 3 | 9 ± 4 | 10 ± 4 | 8 ± 4 | 4 ± 2 | 4 ± 1 |

Five different β phases were expected to form in *M3* mixtures in which LLL, MMM and PPP molecules were included: $β_{LLL}$ ($d \approx 3.15$ nm), $β_{LLL/MMM}$ (3.39 nm), $β_{MMM}$ (3.57 nm), $β_{MMM/PPP}$ (3.78 nm) and $β_{PPP}$ (4.02 nm). The *d*-spacings for the LLL+MMM phases are those determined in the experiments described above, whereas the $d$-spacing for $\beta_{MMM/PPP}$ phase was determined in a separate experiment with binary 1:1 MMM+PPP mixture, see Supporting Figure S13. Indeed, using these 5 phases we were able to perform peak deconvolution analysis and to obtain very high $R^2$-values for both *M3a* (LLL:MMM:PPP = 1:1:1) and *M3b* (LLL:MMM:PPP = 60:25:15) mixtures, see Figure 9c,d and Table 2. The only difference between these samples was the relative peak areas for the different phases: $(\beta_{LLL} + \beta_{LLL/MMM}) : (\beta_{MMM} + \beta_{MMM/PPP} + \beta_{PPP}) \approx$ 21:79 ± 6% for *M3a* and ≈ 51:49 ± 4% for *M3b* samples. We note that the individual peak areas are determined with a relatively big error (see Table 2) because several different deconvolution analyses can be made yielding similar $R^2$ values. Nevertheless, the ratios presented above are determined with better accuracy because the LLL peaks are well distinguished from the MMM/PPP peaks. Note that the total MMM+PPP content in both *M3* samples is smaller when compared to the observed peak areas. Thus, we expect that some fraction of the LLL molecules is included in the mixed $\beta_{MMM/PPP}$ phase, but without changing it characteristic interlamellar spacing.

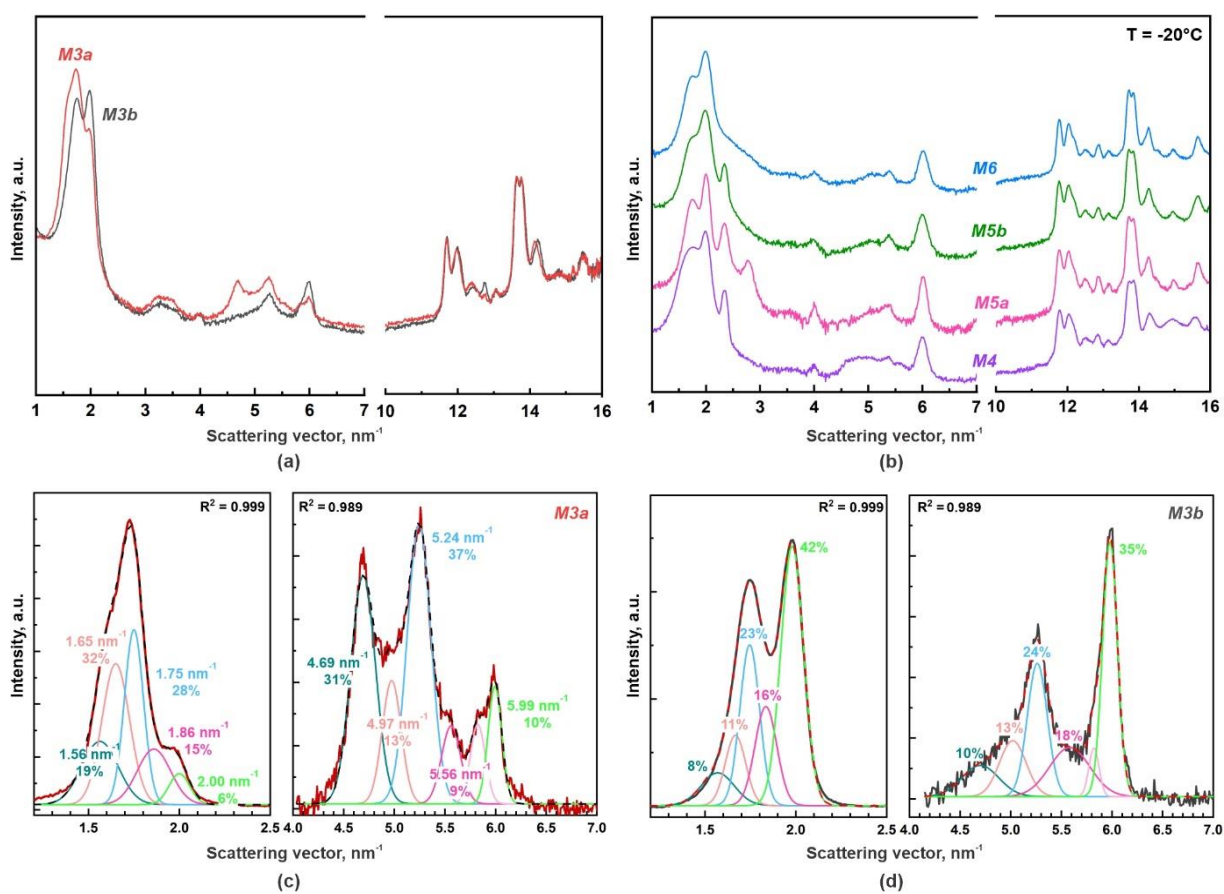

**Figure 9. Beta phases in multicomponent mixtures.** (a) SWAXS spectra for 3-component mixtures. The peak deconvolution analysis for these spectra is presented in (c,d). Peak maxima

and relative peak areas are written on the graphs. (b) SWAXS spectra for 4- and 5-component mixtures. Samples composition is given in Table 1.

The further increase of the number of mixed TAG species increased also the number of phases that can be formed in the most stable polymorphic phase. Seven different β phases were observed in *M4* mixture (CCC:LLL:MMM:PPP = 15:50:20:15), see Table 2. They included the 5 phases already identified in *M3* mixtures (*i.e.* in the absence of CCC) as well as $β_{CCC/LLL}$ and $β_{CCC}$ phases. Illustrative peak deconvolution analysis is presented in Supporting Figure S14.

Figure 9b shows a comparison between the β phases formed in *M5* and *M6* mixtures. To obtain these profiles we first cooled these samples to -20°C at 20°C/min rate, then heated them to 35°C at 1.5°C/min rate to allow for the polymorphic phase transitions to occur, and finally decreased the temperature down to -20°C again to allow for crystallization of all TAGs, because OOO and CaCaCa were in a liquid state at $T \geq 30°C$ (temperatures needed for the LLL rearrangement into β domains). *M5* and *M6* mixtures have similar TAG composition to *M4* sample (see Table 1) and, hence, they crystallized in similar α or β' phases as explained in the previous section. Similar β phases were observed as well upon heating. Once again, these β phases were explained assuming the formation of pure β domains containing different (saturated) TAGs, coexisting with binary mixed β phases, Table 2. The only noticeable difference between the various spectra was that $β_{CCC}$ phase had a well pronounced peak in *M5a* mixture (the pink curve in Figure 9b), whereas this peak was missing when CaCaCa was exchanged by OOO (the green curve in Figure 9b), or it was smaller when both CaCaCa and OOO were present (the blue curve in Figure 9b). We note that separate peaks for OOO could not be identified in *M5b* and *M6* samples because the position of the peak is similar to the one for PPP.

### 3.2.3 *Melting trends*

The DSC thermograms of the quickly cooled *M4*, *M5* and *M6* mixtures (see Table 1 for composition of these mixtures) showed small endothermic peak ($\Delta H \approx$ 3-10 J/g), prior to the polymorphic phase transitions and the subsequent main melting, see the red regions in Figure 7b and Supporting Figure S15. The structural analysis revealed that this peak corresponds to melting of the domains containing the shortest molecules in the mixtures – tricaprylin and/or tricaprin, see Figure 8c. The same process was also observed in the SAXS curves of the slowly cooled samples, see Supporting Figure S11a,b. However, this melting was not visible in the DSC experiments of the slowly cooled samples because it overlapped with the polymorphic phase transition occurring simultaneously with greater enthalpy, see Supporting Figure S15. This result indicates coexistence between a small fraction of already molten molecules and frozen domains of the longer TAG molecules – these crystalline domains remain present up to much higher temperatures. We expect that the distinct crystalline domains are surrounded by the liquid phase of the shorter molecules, similarly to the melting behavior observed with the two-component mixtures, see Figure 3e.

The DSC analysis shows several important trends, regarding the main melting process observed at $T > 25°C$. The melting enthalpy decreases with the complexity of the mixtures. This enthalpy decrease cannot be explained only by considering the fraction of molecules melting at lower temperatures (CCC or CaCaCa), because their contribution into the overall enthalpy of the mixture is relatively small. Furthermore, if one estimates the "theoretical melting enthalpy" using $\Delta H_m$ for the pure triglycerides and their fraction in the mixtures, much higher value of $\Delta H \approx 195 \pm 6$ J/g is expected, instead of the observed experimentally $\Delta H \approx 140$ to 156 J/g for 4-6 component TAG mixtures (see Figure 7). The measured smaller enthalpies evidence that the degree of ordering inside the crystal domains decreases greatly when different chain-lengths are mixed, due to the disturbed packing in the latter systems. Note that even lower enthalpy was measured with CNO, $\Delta H_{m,CNO} \approx 110$ J/g. In addition, the melting temperatures of *M6* and CNO are also quite different, reflecting the difference in the molecular packing – the complete CNO melting is observed at *ca.* 28°C, while the six-component mixture melts completely at $T \approx 55°C$, see Table 3.

The melting temperatures of the other model mixtures were also determined primarily by the longest TAGs in the mixture and varied between *ca.* 52°C for *M2b* (7:3 LLL+MMM) to 62°C for *M3a* which contains 33.3% PPP. The temperature interval of (complete) melting widens significantly when TAGs with a shorter chains and lower melting temperatures are present. For example, the melting interval is *ca.* 10°C for the binary LLL+MMM mixture, whereas it becomes nearly 20°C for the three-component mixtures and increases up to more than 30°C for the six-component mixture when the latter is slowly heated after quick cooling, see the orange regions in Figure 7b.

The melting of the saturated TAGs always started with the TAG having the shortest chains and proceeded consecutively until the melting temperature of the longest TAG in the mixture was approached. However, the melting temperature of the PPP phases in all mixed samples was by *ca.* 4-10°C lower compared to that of the pure bulk PPP ($T_{mPPP} \approx 66°C$), most probably due to the molecular solubility of the PPP molecules in the surrounding, already melted triglyceride phase.

**Table 3. Phase behavior of model triglyceride mixtures upon cooling with 20°C/min or 1.5°C/min rate and subsequent slow heating**. The crystallization temperatures are determined from DSC measurements, whereas the rest of the data is determined from SWAXS experiments. The numbers presented in bold denote the peaks with the highest intensities, see Table 1 for the composition of the studied TAG mixtures.

| | | Model triglyceride mixtures | | | | | | | Coconut oil |
|---|---|---|---|---|---|---|---|---|---|
| | | *M2a* | *M3a* | *M3b* | *M4* | *M5a* | *M5b* | *M6* | |
| Cooling 20°C/min | $T_{cr}$ start, °C | 22.7 | 30.5 | 24.2 | 24.7 | 20.1 | 22.1 | 19.6 | 13.2 |
| | $T_{cr}$ peak, °C | 18.9<br>**13.3** | **27.4**<br>10.7 | 19.7<br>**9.6** | 21.5<br>**7.1** | 15.7<br>**4.7** | 18.7<br>**6.8** | 16.7<br>**6.2** | 2.6<br>**-11.5** |
| | Polymorphs | α | α + β' | α + β' | α + β' | α + β' | α + β' | α + β' | α + β$_1$' |
| | *d*, nm | **3.99**<br>3.55 | 4.42<br>**4.20**<br>3.67<br>*3.38* | 4.31<br>3.99<br>**3.57**<br>3.35 | 4.38<br>4.05<br>**3.32**<br>2.97 | 4.32<br>3.97<br>3.62<br>**3.33**<br>3.04 | 4.31<br>4.03<br>3.63<br>**3.32**<br>3.04 | 4.36<br>4.12<br>3.55<br>**3.31**<br>3.14 | **3.82**<br>3.30 |
| Heating 1.5°C/min | *T* β only, °C | 31 | 40 | 40 | 33 | 24 | 24 | 10 | 10* |
| | $T_m$ end, °C | 58 | 62 | 58 | 57 | 54 | 54 | 55 | 28 |
| Cooling 1.5°C/min | $T_{cr}$ start, °C | 32.8 | 33.5 | 30.3 | 31.9 | 29.5 | 31.5 | 28.5 | 16.0 |
| | $T_{cr}$ peak, °C | **28.1** | 32.0<br>**30.3**<br>26.7 | 25.1<br>23.3 | **25.2**<br>9.5 | 22.0<br>4.1 | 21.7<br>5.9 | **19.2** | 7.0 |
| | Polymorphs | β' | β' | β' | β' | β' + β | β' + β | β' + β | β$_2$' |
| | *d*, nm | **3.57**<br>3.41 | 3.98<br>**3.76**<br>3.35 | **3.69**<br>3.41<br>3.30 | 3.88<br>**3.60**<br>3.40<br>**3.29**<br>2.91 | 3.93<br>**3.64**<br>3.52<br>3.37<br>3.27<br>3.16<br>2.84<br>2.39 | 4.14<br>**3.64**<br>**3.29**<br>3.13<br>2.93<br>2.67 | 4.12<br>3.70<br>3.34<br>**3.15**<br>2.70<br>2.24 | 3.39 |
| Heating 1.5°C/min | *T* β only, °C | 37 | 42 | 37 | 34 | 25 | 22 | 22 | - |
| | $T_m$ end, °C | 56 | 63 | 57 | 58 | 55 | 55 | 55 | 28 |

*For CNO, the β$_2$' phase is thermodynamically stable, instead of β phase [39].

## 4. Conclusions

We studied the polymorphic phase behavior of various TAG mixtures, upon their cooling and subsequent heating. The experimental conditions, such as mixture composition, applied cooling rate, storage time and temperature, were varied systematically in wide ranges. The main observed trends are briefly summarized below:

Upon rapid cooling with rate ≥ 30°C/min, two immiscible α phases are formed in the binary LLL+MMM mixtures. Each of these α phases is a mixture of LLL and MMM molecules of different ratios and their interlamellar spacing increases with the MMM fraction in the overall mixture. Upon rapid cooling of three-component or more complex in composition mixtures, three or more mixed immiscible phases are formed which could be of α and β' type.

Upon slow cooling of 1°C/min, the TAG molecules in the two- and three-component mixtures arrange in mixed β' domains, whereas several β' and β domains are formed in the mixtures containing four or larger number of components.

Upon slow heating with *ca.* 1°C/min, these α and β' phases transform into the most stable β phases. The structure of the formed β polymorphs does not depend on the thermal history and on the specific composition of the mixture. As a rule, $(2k - 1)$ different β phases are formed in the mixtures containing $k$ different even-numbered saturated TAG species: $k$ single-component β phases for all individual TAG species in the mixture, along with $(k-1)$ binary mixed β phases, composed of TAGs with neighboring chain-lengths. For example, $β_{LLL}$, $β_{LLL/MMM}$ and $β_{MMM}$ phases were observed in all LLL+MMM mixtures. For all TAGs studied ($C_8$ to $C_{16}$ fatty acid residues) and their binary mixtures, the interlamellar spacing can be estimated as $d_β ≈ 0.222\bar{n} + 0.4682$ nm, where $\bar{n}$ is the average number of C-atoms in the fatty acid residues, e.g. $\bar{n} = 13$ for the LLL+MMM mixture.

The actual melting of the TAG mixtures upon heating starts from the shortest molecular species with the lowest melting temperature. Coexistence between the already molten TAG species and still frozen domains is observed in wide temperature intervals, when TAGs with significantly different chain-lengths are mixed.

The current study advances significantly our understanding of the polymorphism in mixtures of monoacid TAGs which are often used as model systems for the natural oils and fats. Such knowledge is particularly important in several application areas, *incl.* foods, beverages, cosmetics and pharmaceuticals. Our results can serve also as a basis for other studies in related areas, *e.g.* to clarify the effects of various additives (medical drugs, essential nutrients, solid nanoparticles, *etc.*) and/or confinement (bulk *vs* layers *vs* microencapsulated/emulsified samples) on the polymorphic transitions.

**Supporting Information.** The Supporting Information is available free of charge on the ACS Publications website. SWAXS and DSC curves for LLL+MMM mixtures; SWAXS curves for *M3a*, *M3b*, *M4*, *M5a*, *M5b* and *M6* TAG mixtures, and for Coconut oil; β phases in MMM+PPP = 1:1 mixture; DSC curves for *M5a* and *M6* TAG mixtures (PDF).

**Acknowledgements:**

The authors thank Ms. Desislava Glushkova (Sofia University) for performing part of the experiments and for useful discussions. We thank the anonymous Reviewer 3, whose valuable suggestions helped us to improve our data interpretation for the β polymorphs. The study was funded by Bulgarian Ministry of Education and Science, under the National Research Program "VIHREN", project ROTA-Active (no. KP-06-DV-4/16.12.2019). The authors acknowledge the possibility to use SAXS/WAXS instrument purchased for execution of project BG05M2OP001-1.002-0012, Operational Program "Science and Education for Smart Growth", Bulgaria.

**CRediT authorship contribution statement:**

**Diana Cholakova:** conceptualization, methodology, investigation, validation, formal analysis, visualization, writing – original draft, review & editing.

**Slavka Tcholakova:** conceptualization, supervision, writing – review & editing, funding acquisition

**Nikolai Denkov:** conceptualization, supervision, writing – review & editing, funding acquisition

# Supporting Information

# Polymorphic phase transitions in bulk triglyceride mixtures

**Diana Cholakova, Slavka Tcholakova, Nikolai Denkov***

*Department of Chemical and Pharmaceutical Engineering*
*Faculty of Chemistry and Pharmacy, Sofia University,*
*1 James Bourchier Avenue, 1164 Sofia, Bulgaria*

*Corresponding authors:
Prof. Nikolai Denkov
Department of Chemical and Pharmaceutical Engineering
Sofia University
1 James Bourchier Ave.,
Sofia 1164
Bulgaria
E-mail: nd@lcpe.uni-sofia.bg
Tel: +359 2 8161639
Fax: +359 2 9625643

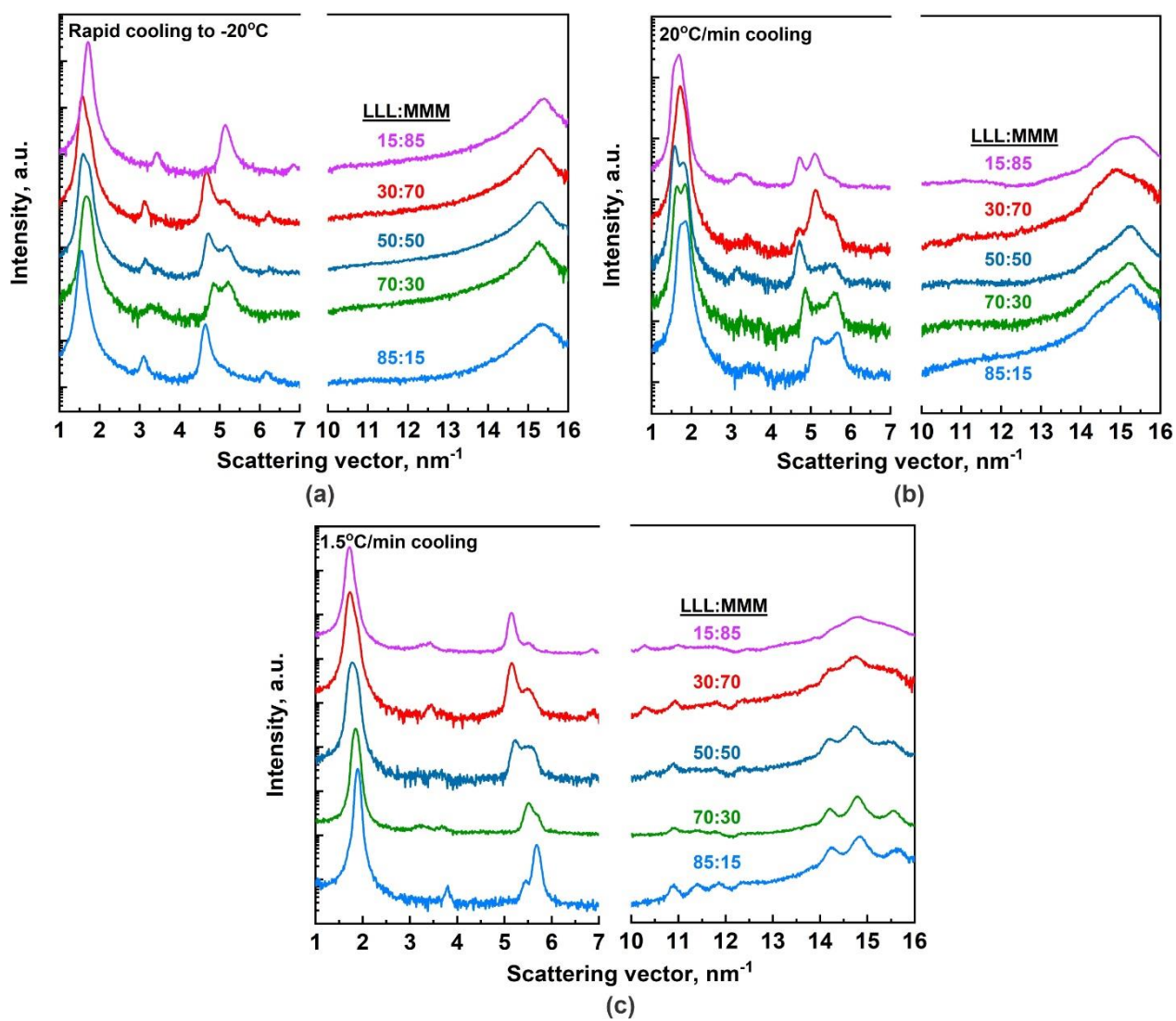

**Supporting Figure S1.** **SWAXS curves obtained at different cooling rates for LLL+MMM mixtures.** **(a)** Rapid cooling to -20°C from melt. α-polymorphs form in the samples. **(b)** Controlled cooling with 20°C/min rate. Two or even three α + β' co-existing phases form in the samples. **(c)** Controlled cooling with 1.5°C/min rate. Samples with ≥ 50 wt. % crystallize in two co-existing β' polymorphs, whereas in the samples with predominant MMM content, both α and β' domains are formed. The phase identification is based on the WAXS spectra.

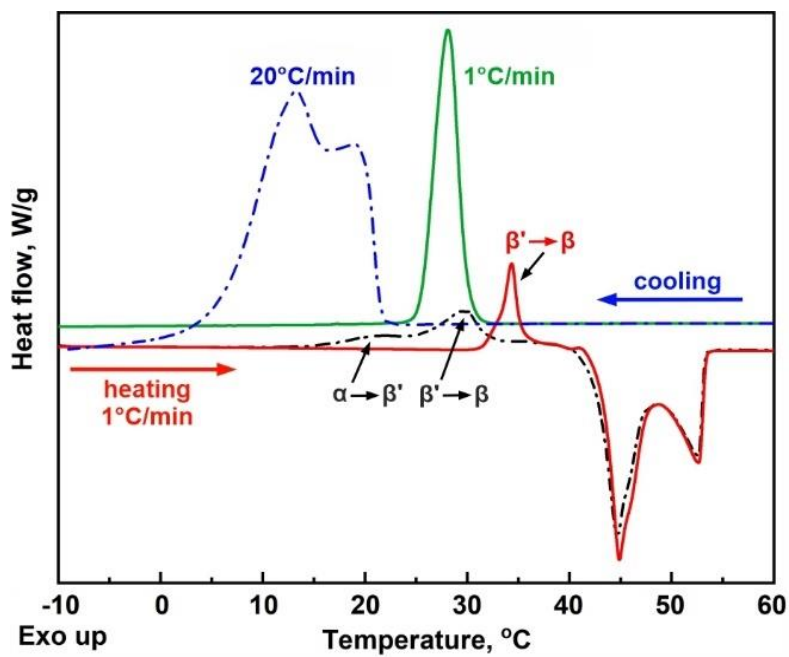

**Supporting Figure S2.** **DSC curves obtained upon cooling and heating of *M2a* sample (LLL+MMM = 1:1).** The cooling rate determines the crystallization temperature and polymorphic phase formed upon freezing: α domains are formed upon 20°C/min cooling, whereas β' domains form upon slower cooling. Upon heating, the samples undergo polymorphic phase transition(s) forming the most stable β polymorph. The melting of β domains does not depend on the temperature history of the sample. In both cases – initially LLL molecules melt, followed by the melting of the longer MMM molecules at slightly higher temperature (see the double maxima).

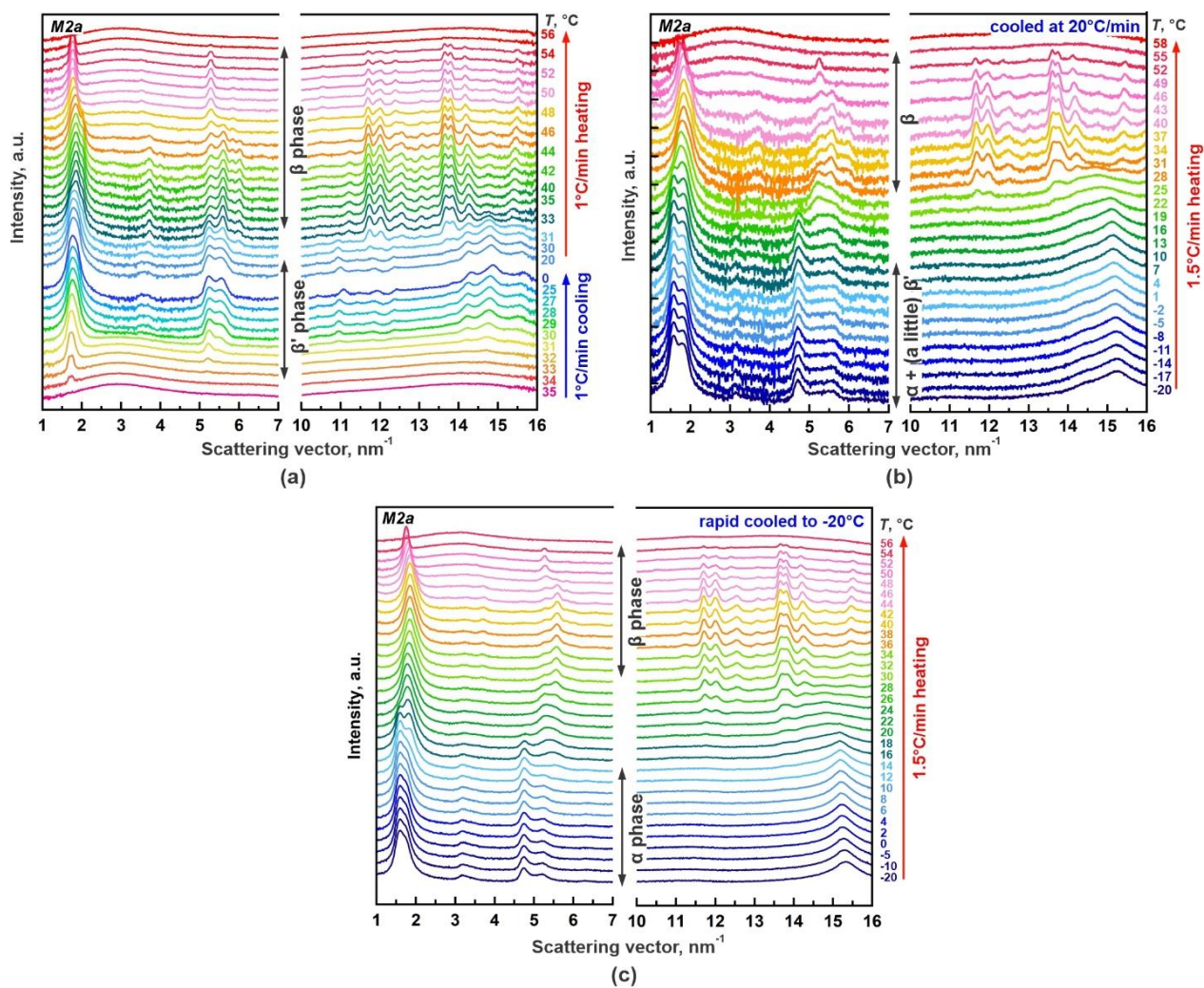

**Supporting Figure S3.** SWAXS curves obtained for *M2a* mixture (LLL + MMM = 1:1) under different cooling-heating protocols. (a) 1°C/min cooling and subsequent heating. (b) 1.5°C/min heating of sample cooled at 20°C/min rate. (c) 1.5°C/min heating of sample cooled rapid at -20°C.

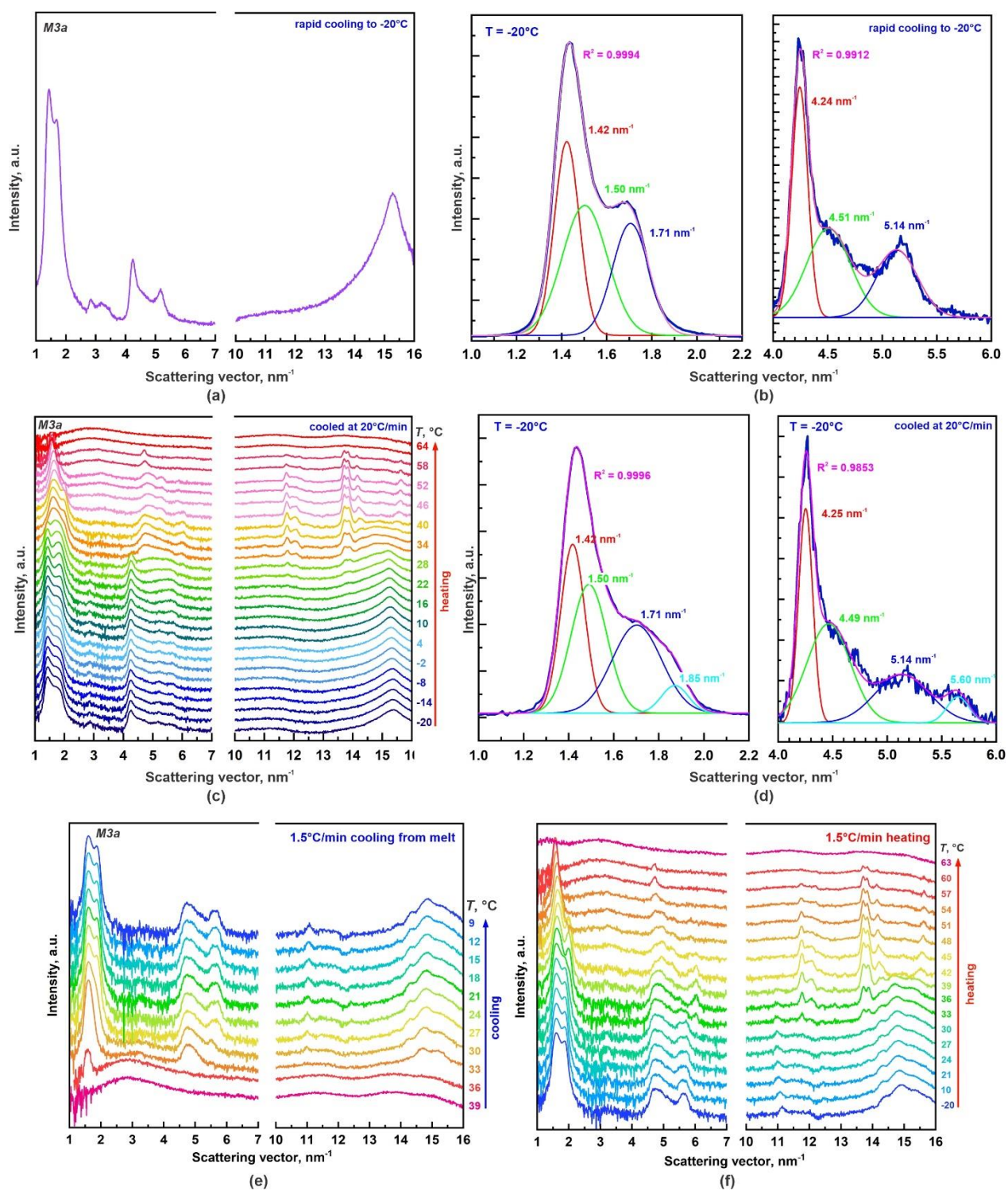

**Supporting Figure S4.** **SWAXS curves for *M3a* mixture (LLL+MMM+PPP = 1:1:1).** (a,b) Scattering curve obtained by rapid cooling to -20°C and the respective peak deconvolution analysis. (c,d) Scattering curves obtained upon 1.5°C/min heating of quickly cooled sample (cooling rate 20°C/min). Four co-existing phases form in the sample with predominant content of three α phases and very small content of β' phase (the cyan peak). (e,f) Scattering curves obtained upon 1.5°C/min cooling from melt (c) and subsequent heating with 1.5°C/min. Two co-existing β' phases are formed upon cooling in this protocol with characteristic lamellar thicknesses of 3.39 and 3.90 nm, respectively.

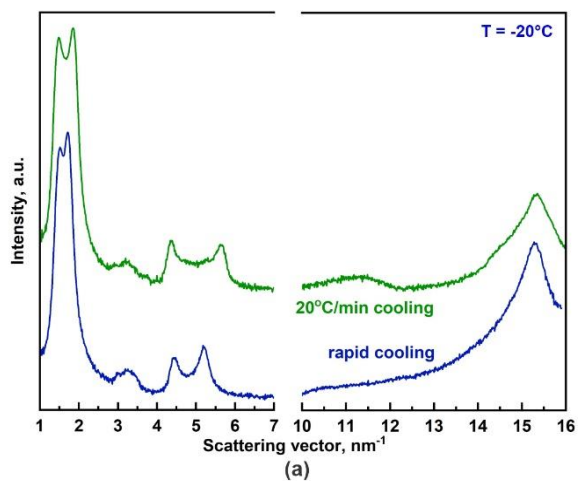
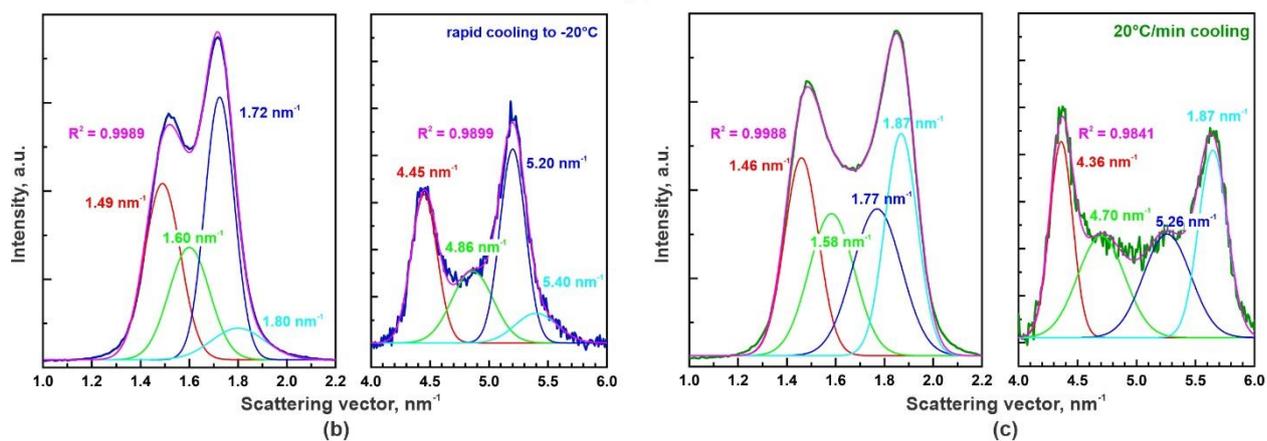

**Supporting Figure S5.** SWAXS curves obtained upon rapid cooling to -20°C (a,b) and at 20°C/min cooling (a,c) of *M3b* mixture (LLL+MMM+PPP = 60:25:15). (b,c) presents peak deconvolution analysis for the spectra obtained at -20°C.

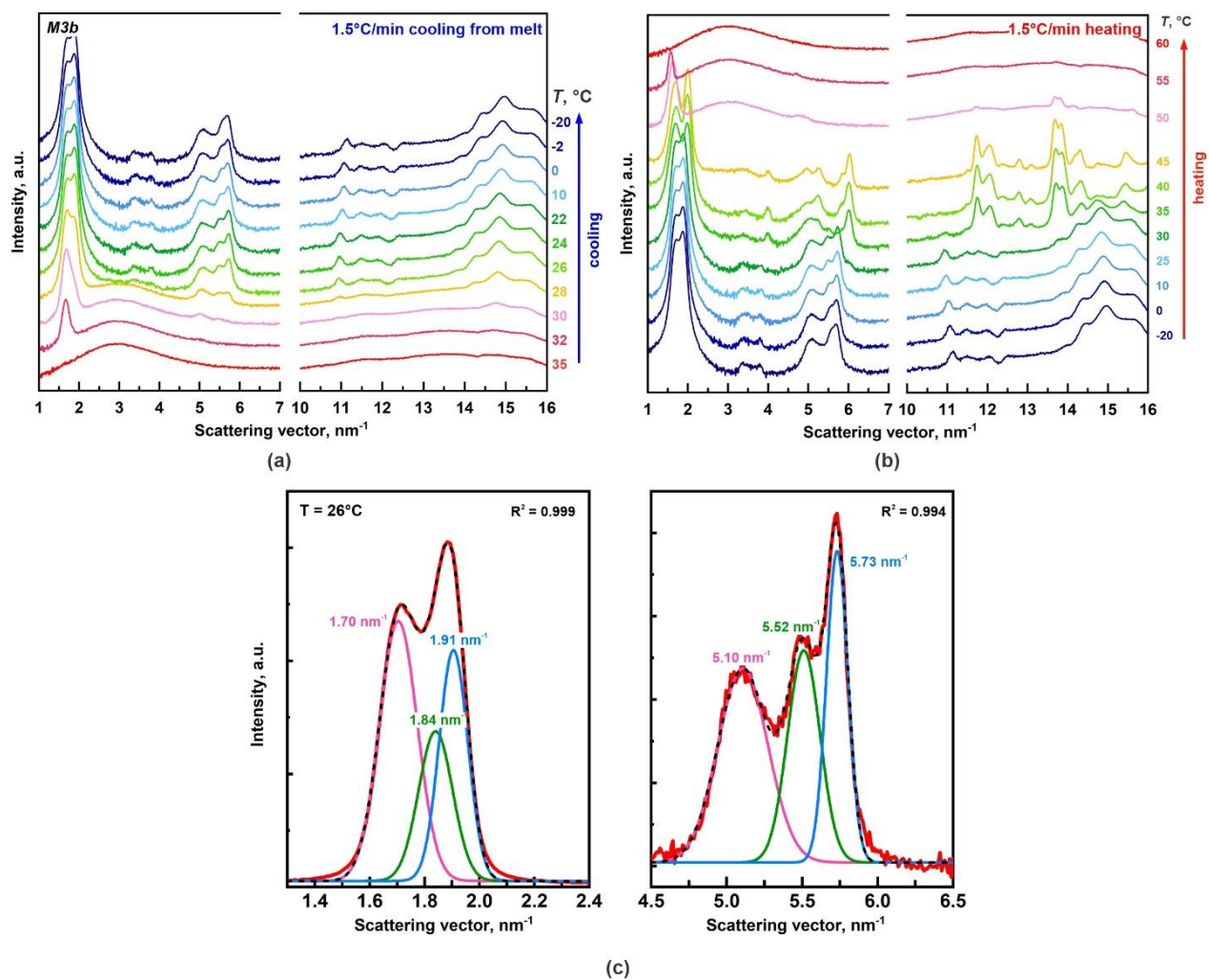

**Supporting Figure S6.** **SWAXS curves obtained upon (a) 1.5°C/min cooling and (b) heating of *M3b* mixture (LLL + MMM + PPP = 60:25:15).** The characteristic lamellar thicknesses of the β' phases formed in this case are 3.30, 3.41 and 3.69 nm, see the peak deconvolution shown in (c).

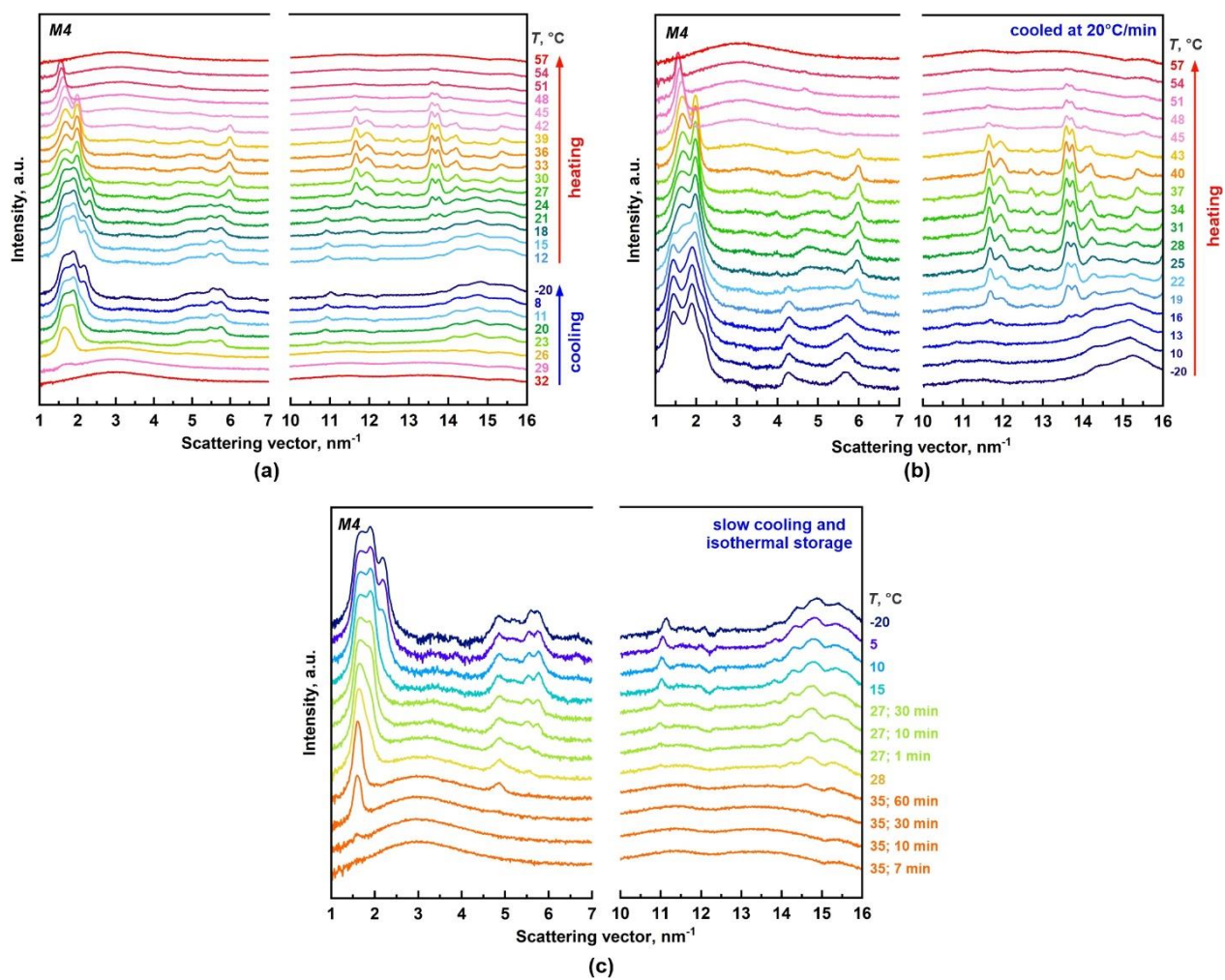

**Supporting Figure S7.** SWAXS curves for *M4* mixture (**CCC+LLL+MMM+PPP = 15:50:20:15**) obtained at different temperature profiles: **(a)** Slow cooling and slow heating at 1.5°C/min. **(b)** Slow heating after fast cooling at 20°C/min. **(c)** Curves obtained upon cooling from 70°C to 35°C at 1.5°C/min cooling rate. After that the temperature is kept constant for 1 hour and the scattering signal is monitored at every few minutes. Note that only one phase appears with $d \approx 3.89$ nm at this temperature. Another four phases appear consecutively upon further cooling and five distinct phases (in total) are observed at -20°C.

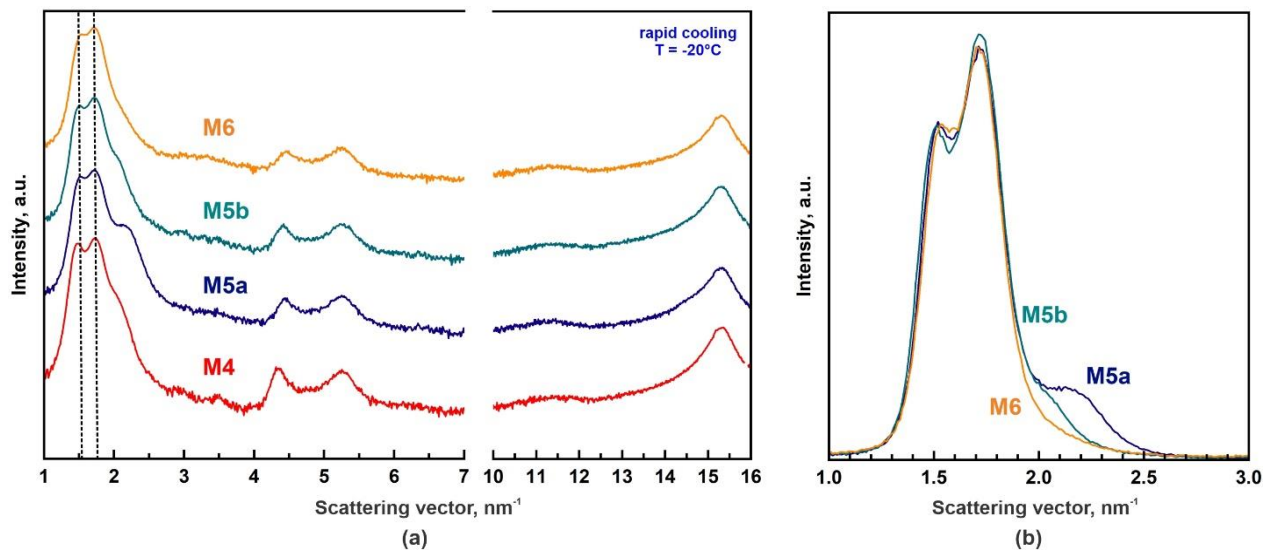

**Supporting Figure S8. SWAXS curves for *M4, M5a, M5b* and *M6* mixtures obtained upon rapid cooling from melt.** All mixtures crystallize in α mixed domains. Small differences are observed only in the phases with the shortest lamellar spacings (b). Mixtures compositions: *M4*: CCC+LLL+MMM+PPP = 15:50:20:15; *M5a*: CaCaCa+CCC+LLL+MMM+PPP = 7:15:50:18:10; *M5b*: CCC+LLL+MMM+PPP+OOO = 15:50:18:10:7 and *M6*: CaCaCa+CCC+LLL+MMM+PPP+OOO = 7:8:50:18:10:7. Note that y-axis is plotted in log scale in (a) while it is linear in (b).

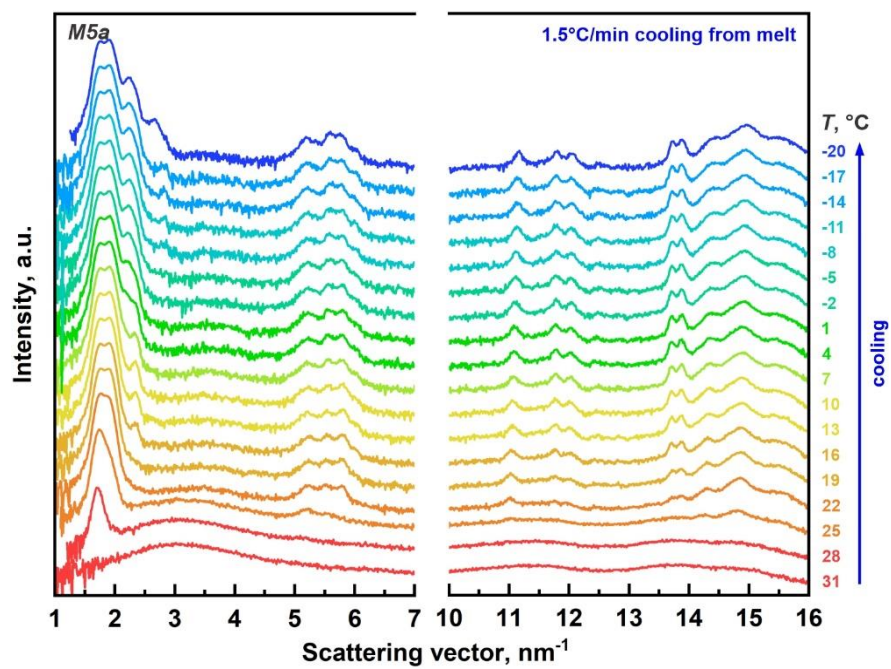

**Supporting Figure S9.** SWAXS curves obtained upon 1.5°C/min cooling of *M5a* mixture (**CaCaCa+CCC+LLL+MMM+PPP = 7:15:50:18:10**). Four co-existing phases are observed with interplanar thicknesses of 4.24, 3.29, 2.91 and 2.31 nm, respectively.

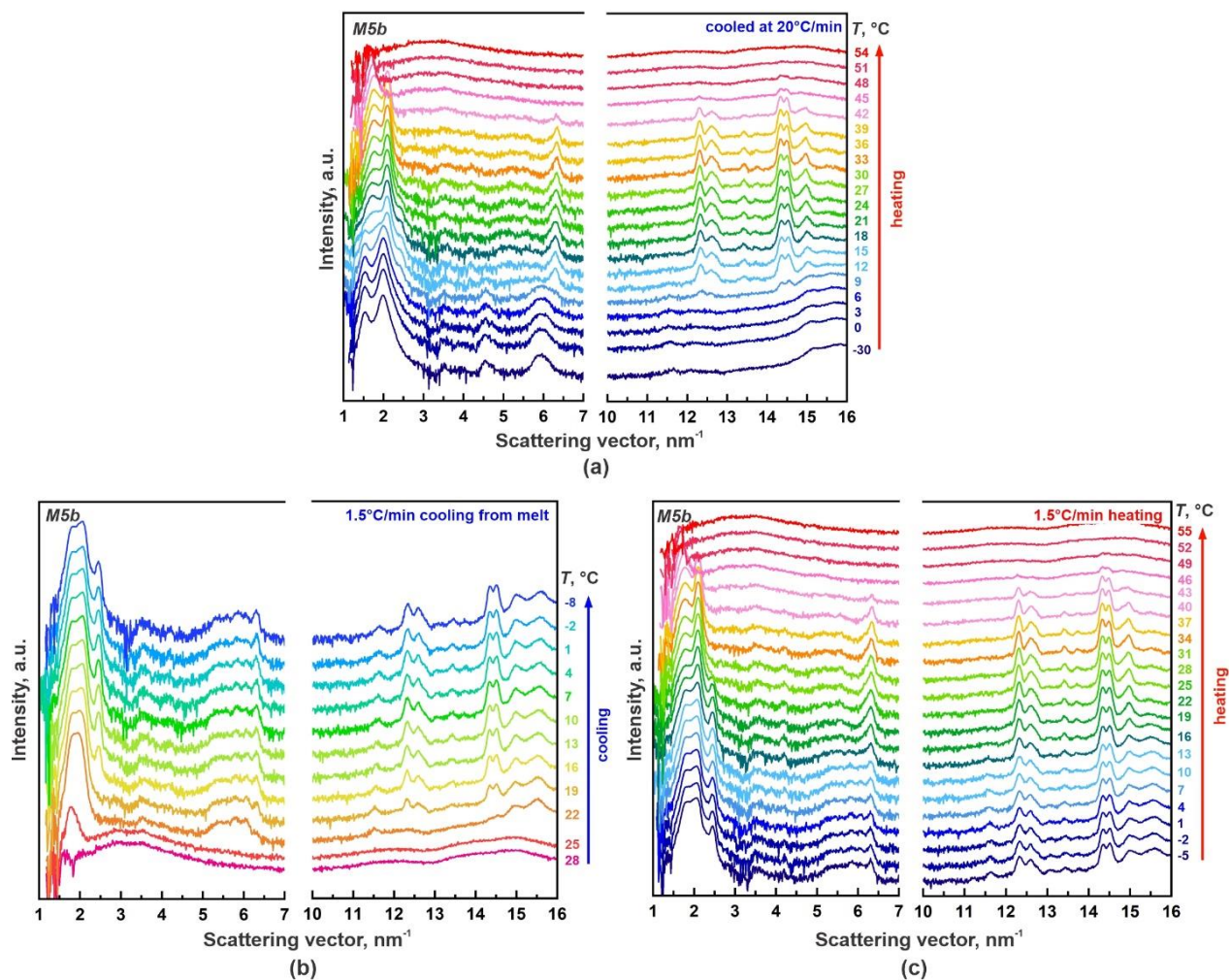

**Supporting Figure S10.** SWAXS curves for *M5b* mixture (CCC+LLL+MMM+PPP+OOO = 15:50:18:10:7). (a) 1.5°C/min heating after cooling at 20°C/min rate. (b,c) 1.5°C/min cooling and subsequent heating. Note the significant difference between the phases obtained with these two cooling protocols – compare the dark blue curves in (a) and (c).

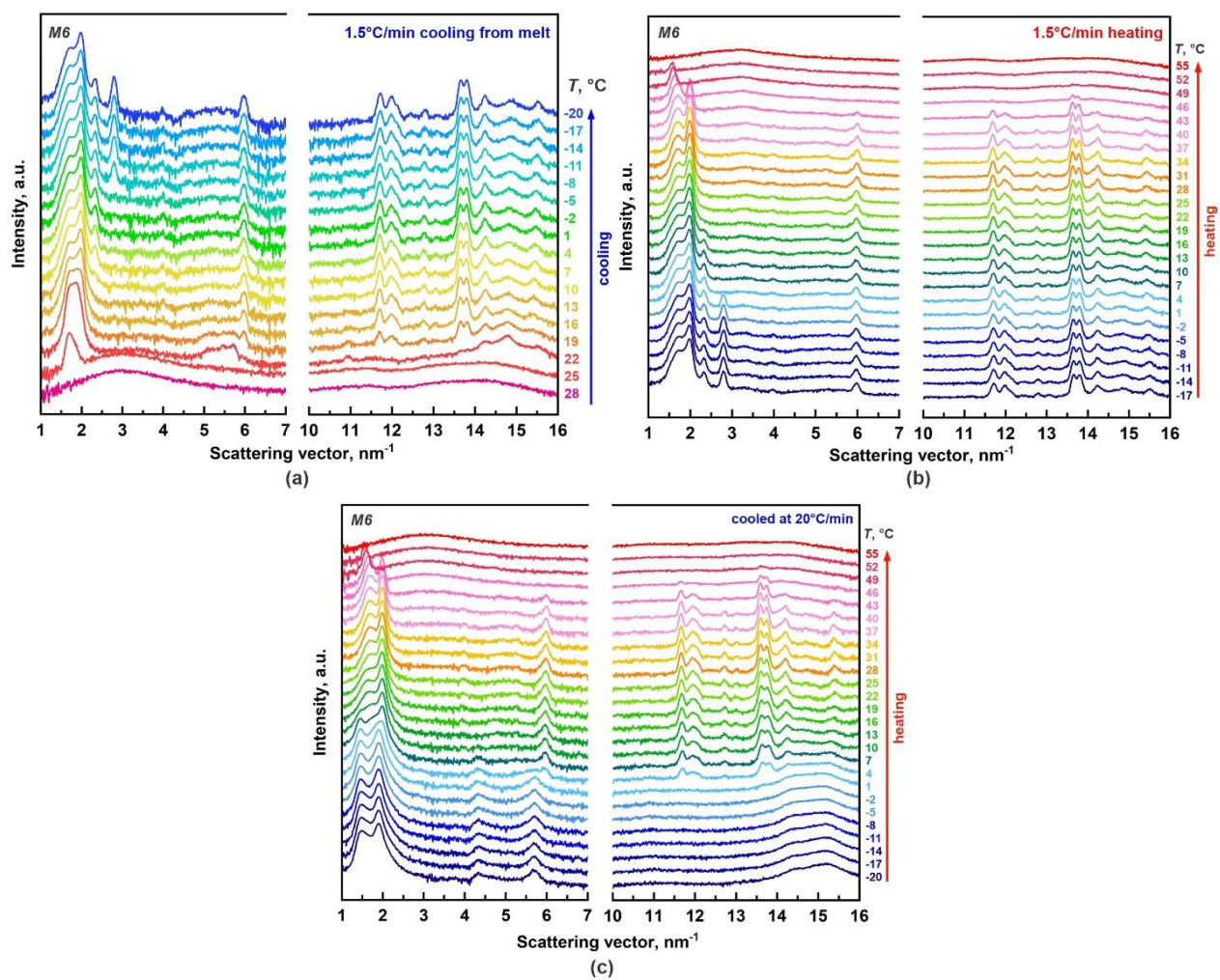

**Supporting Figure S11.** SWAXS curves for *M6* mixture (CaCaCa+CCC+LLL+MMM +PPP+OOO = 7:8:50:18:10:7). (a,b) 1.5°C/min cooling and subsequent heating. (c) 1.5°C/min heating after cooling with 20°C/min rate. Note the significant difference between the phases obtained with these two cooling protocols – compare the dark blue curves in (b) and (c).

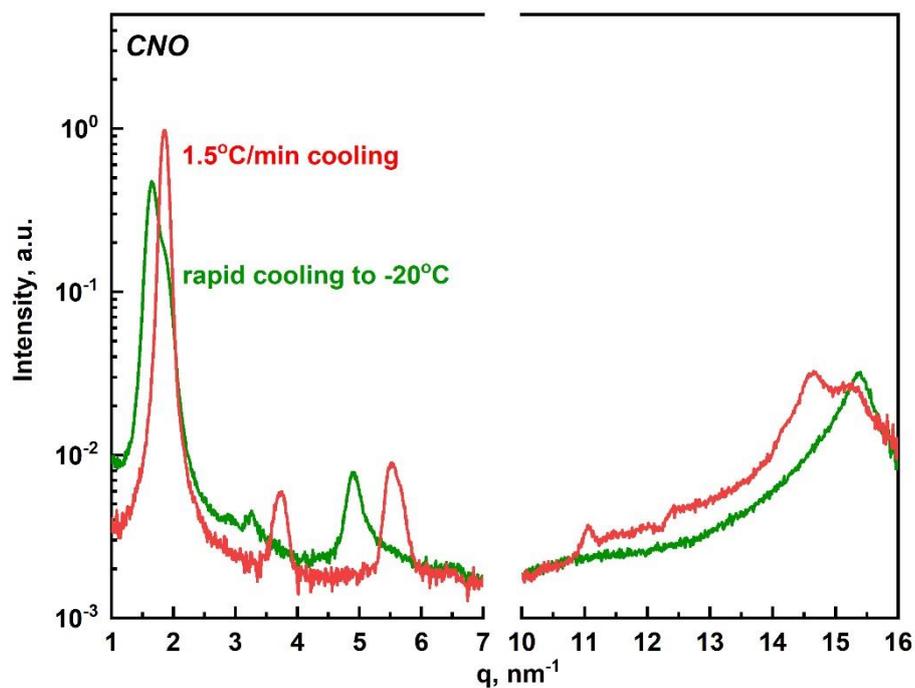

**Supporting Figure S12. SWAXS curves for *CNO*.** Both scattering profiles are obtained at temperature of -20°C, but CNO crystallize in different polymorphs, depending on the cooling rate: α-phase is observed upon direct cooling to -20°C from melt (green curve), whereas $β_2'$ phase is detected upon slower cooling at 1.5°C/min (red curve).

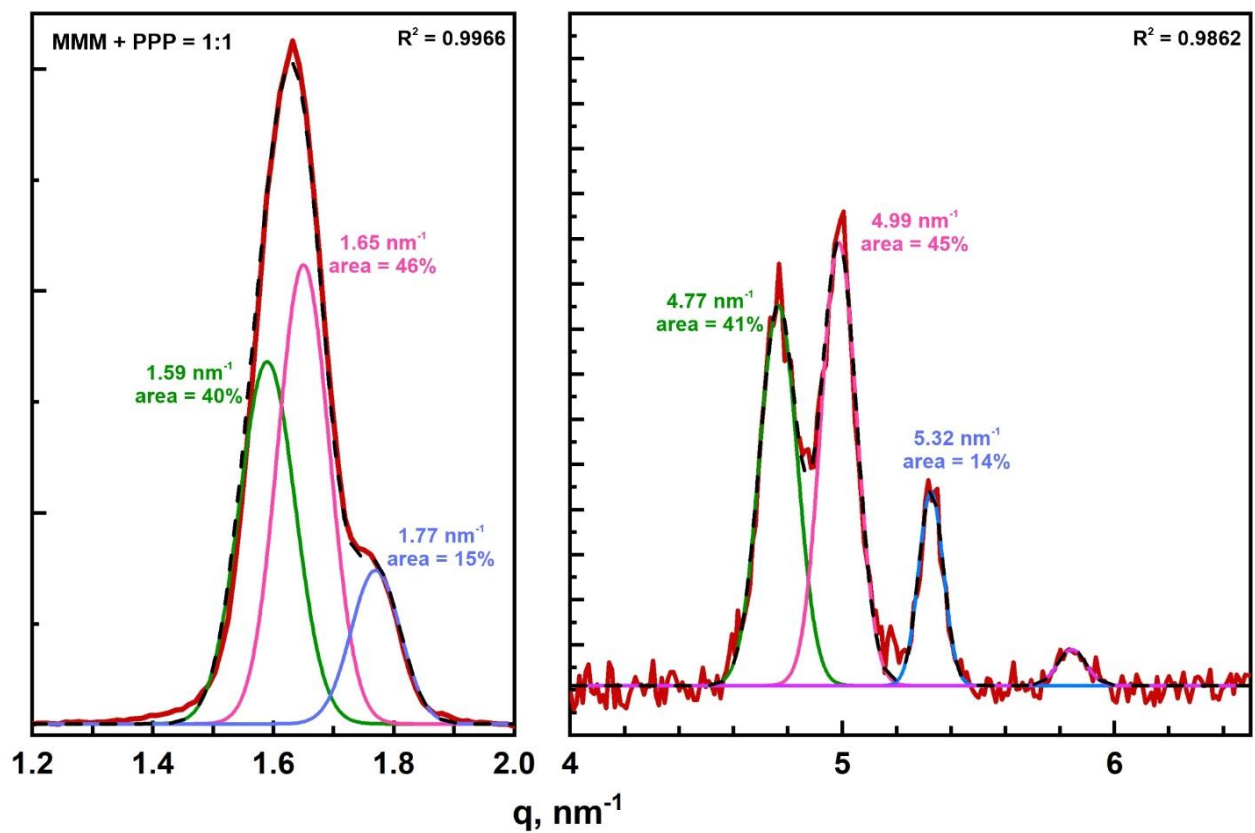

**Supporting Figure S13. β phases observed in MMM+PPP = 1:1 mixture.** Three distinct phases are identified: $β_{PPP}$ ($d ≈ 3.95$ nm), $β_{PPP/MMM}$ ($d ≈ 3.79$ nm) and $β_{MMM}$ ($d ≈ 3.54$ nm).

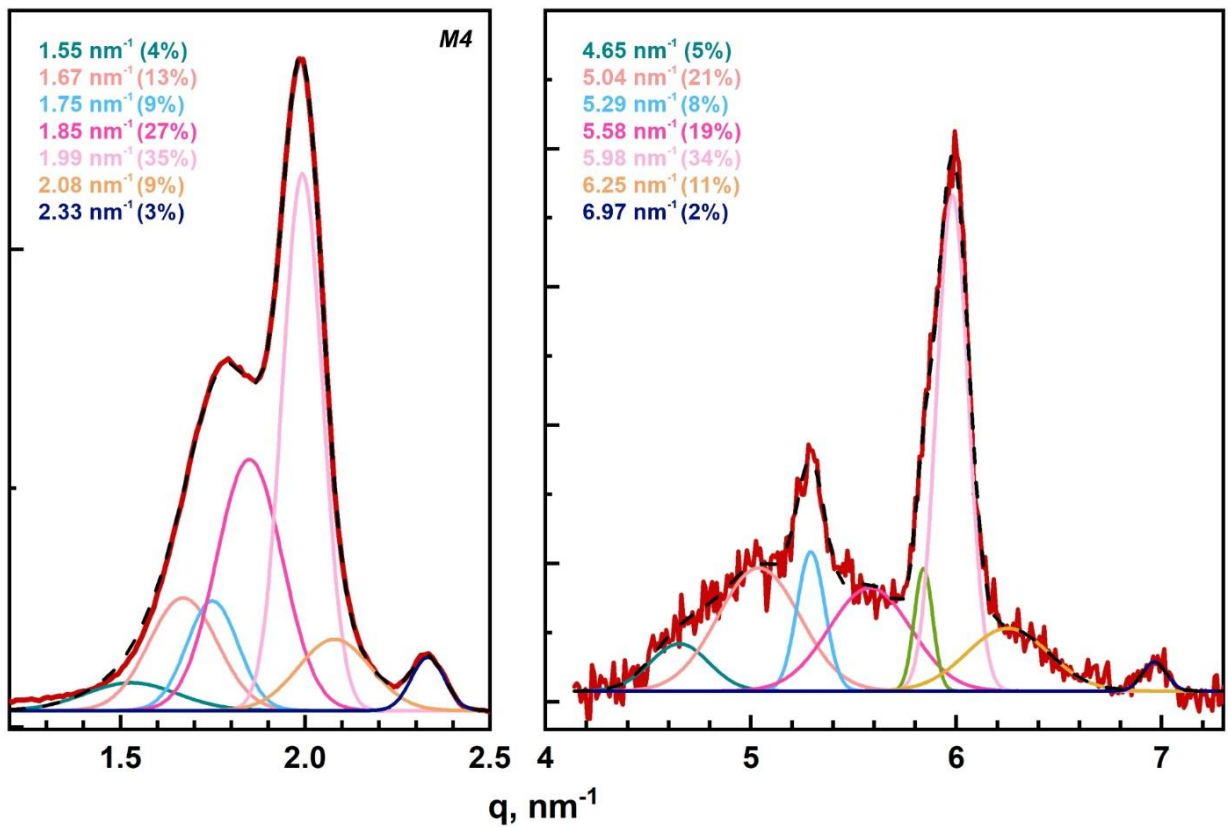

**Supporting Figure S14. Peak deconvolution analysis for β phases observed in *M4* (CCC +LLL+MMM+PPP = 15:50:20:15) mixture.** Seven different types of domains have been identified: PPP, PPP/MMM, MMM, MMM/LLL, LLL, LLL/CCC, and CCC. Peak maxima and relative peak area are denoted on the graphs. Note that the green peak ($q \approx 5.84$ nm$^{-1}$) in the right-hand-side graph is caused by mixed (*hkl*) reflection and it has been disregarded in the analysis.

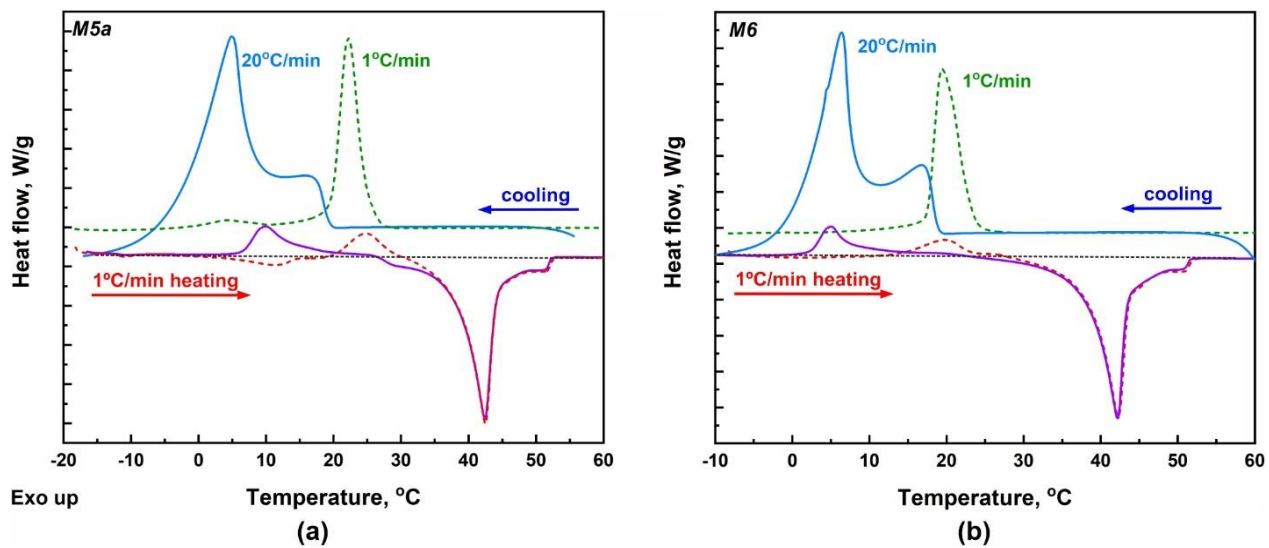

**Supporting Figure S15.** DSC curves obtained upon cooling and heating of: (a) *M5a* and (b) *M6* triglyceride mixtures.